\documentclass{article}
\usepackage[utf8]{inputenc}
\usepackage{amsmath}
\usepackage{amssymb}
\usepackage{amsfonts}
\usepackage{esvect}
\usepackage{cite} 
\usepackage{physics}
\usepackage{todonotes}
\usepackage{mathrsfs}
\usepackage{authblk}
\usepackage{geometry}
\usepackage{abstract}
\usepackage{xcolor}
\usepackage[colorlinks=true,urlcolor=blue]{hyperref}

\numberwithin{equation}{section}

\begin{document}

\thispagestyle{empty}
\begin{center}
	\Large{{\bf Polynomial algebras from Lie algebra reduction chains $\mathfrak{g} \supset \mathfrak{g}'$}}
\end{center}
\vskip 0.5cm
\begin{center}
	\textsc{Rutwig Campoamor-Stursberg$^{1,\star}$, Danilo Latini$^{2,*}$, Ian Marquette$^{2,\dagger}$ and Yao-Zhong Zhang$^{2,\ddagger}$}
\end{center}
\begin{center}
	$^1$ Instituto de Matem\'{a}tica Interdisciplinar and Dpto. Geometr\'{i}a y Topolog\'{i}a, UCM, E-28040 Madrid, Spain
\end{center}
\begin{center}
	$^2$ School of Mathematics and Physics, The University of Queensland, Brisbane, QLD 4072, Australia
\end{center}
\begin{center}
	\footnotesize{$^\star$\textsf{rutwig@ucm.es} \hskip 0.25cm$^*$\textsf{d.latini@uq.edu.au} \hskip 0.25cm $^\dagger$\textsf{i.marquette@uq.edu.au} \hskip 0.25cm $^\ddagger$\textsf{yzz@maths.uq.edu.au}}
\end{center}
\vskip  1cm
\hrule
\begin{abstract}
\noindent  We reexamine different examples of reduction chains $\mathfrak{g} \supset \mathfrak{g}'$ of Lie algebras in order to show how the polynomials determining the commutant with respect to the subalgebra $\mathfrak{g}'$ leads to polynomial deformations of Lie algebras. These polynomial algebras have already been observed in various contexts, such as in the framework of superintegrable systems. Two relevant chains extensively studied in Nuclear Physics, namely the Elliott chain $\mathfrak{su}(3) \supset \mathfrak{so}(3)$ and the chain $\mathfrak{so}(5) \supset \mathfrak{su}(2) \times \mathfrak{u}(1)$ related to the Seniority model, are analyzed in detail from this perspective. We show that these two chains both lead to three-generator cubic polynomial algebras, a result that paves the way for a more systematic investigation of nuclear models in relation to polynomial structures arising from reduction chains. In order to show that the procedure is not restricted to semisimple algebras, we also study the chain $\hat{S}(3) \supset \mathfrak{sl}(2,\mathbb{R}) \times \mathfrak{so}(2)$ involving the centrally-extended Schr{\"o}dinger algebra in $(3+1)$-dimensional space-time. The approach chosen to construct these polynomial algebras is based on the use of the Lie-Poisson bracket, so that all the results are presented in the Poisson (commutative) setting. The advantage of considering this approach {\it vs.} the enveloping algebras is emphasized, commenting on the main formal differences between the polynomial Poisson algebras and their noncommutative analogue. As an illustrative example of the latter, the three-generator cubic algebra associated to the Elliott chain is reformulated in the Lie algebraic (noncommutative) setting. 
\end{abstract}
\vskip 0.35cm
\hrule

\section{Introduction}
\label{int}

Finitely generated polynomial algebras, namely polynomial deformations of Lie algebras, have found to appear in various problems of mathematical physics, such as in the interpretation of the symmetries of Racah coefficients ($6j$-symbols) and their use in the Clebsch-Gordan problem \cite{gra88, gra93}. They have also been recognized to be a powerful tool in the analysis of superintegrable systems (i.e. finite-dimensional Hamiltonian dynamical systems characterized by a number of constants of the motion exceeding the number of degrees of freedom) \cite{zhe93, bon94,Deb0,Deb}, where these structures and their Casimir invariants have been used to determine the energy spectra of Hamiltonians \cite{Hig,das01,Stan,Civ}. For instance, quadratic, cubic and higher-order polynomial deformations of Lie algebras have played a significant role in classification of superintegrable systems, and have deep connections with the theory of orthogonal polynomials and generalized special functions \cite{Kre07, mar09, mar092, mar10,pos11, mil13, kal13}. Some types of higher-rank quadratic algebras have also been used to solve $n$-dimensional Schr{\"o}dinger equations by pure algebraic methods \cite{lia18, lat19, lat21, lat21ch, cor21sw} . It is worthy to be mentioned that these algebraic structures are also connected with several other mathematical approaches, like PBW algebras or Gr\"obner bases \cite{pol11,li11,jar11,li21,yat18, mar22}, from which applications to signal analysis have been deduced \cite{gru18}.

\medskip
In the recent paper \cite{cra21} a polynomial cubic algebra whose defining relations were initially discovered in \cite{RacahLect} (without explicit formulas for the coefficients)  was discussed in the framework of the so-called missing label problem associated to the chain $SU(3) \supset SO(3)$ \cite{mos63, jud74, mpsw, plu86}.  These cubic relations were obtained in terms of two missing label operators, of degree-three and degree-four respectively, defined in the enveloping algebra $\mathcal{U}(\mathfrak{su}(3))$.
We also mention that these cubic relations are in fact a particular case of a more general cubic algebra studied in \cite{cra20}. These reduction chains, and associated missing label operators, often appear in connection with dynamical models in nuclear physics, where the algebraic Hamiltonian is expressed as a function of the invariant operators of the chain. In this context, we mention the Elliott model based on the chain $\mathfrak{su}(3) \supset \mathfrak{so}(3)$ \cite{ell1, ell2},  the interacting Boson model based on $\mathfrak{u}(6)$ \cite{ia87} as well as the nuclear quadrupole model based on the chain $\mathfrak{so}(5) \supset \mathfrak{so}(3)$ \cite{ker, bhh07}. 
Finding missing labels associated to non-canonical chains is in general a very difficult problem \cite{iacbook}, and many different approaches have been developed to deal with the problem, with the infinitesimal approach being the most common method used to construct the generalized Casimir invariants of Lie algebras, as well as the labelling operators and the associated bases of eigenstates \cite{Pe76,Cheb,swbook}.  Within the analytical approach, the Lie algebra generators are realized in the space  $C^\infty(\mathfrak{g}^*)$ by means of the coadjoint representation (summation over repeated indices is understood):
\begin{equation}\label{difo1}
\hat{X}_i = C_{ij}^k x_k \partial_{x_j} \, ,
\end{equation}  
where $\boldsymbol{x}:=(x_1, \dots, x_d)$ represent the corresponding (commuting) coordinates in the dual vector space $\mathfrak{g}^*$ of  the Lie algebra $\mathfrak{g}$, and $d=\text{dim}(\mathfrak{g})$. The problem of finding invariants is then restricted to find the solutions of the following system composed of $d$ linear Partial Differential Equations (PDEs):
\begin{equation}
\hat{X}_i(f(\boldsymbol{x}))=C_{i j}^k x_k \frac{\partial f(\boldsymbol{x})}{\partial x_{j}} =0, \quad i=1, \dots, d=\dim(\mathfrak{g}) \label{eq:casinv}
\end{equation}
whose maximal number of functionally independent solutions is known to be \cite{BB, Pau}:
\begin{equation}\label{difo2}
N(\mathfrak g)= \text{dim} (\mathfrak g)- \text{rank} ||C_{ij}^k x_k||, \quad i,j,k = 1, \dots, d,
\end{equation}
where $C_{ij}^k$ are the structure constants of the Lie algebra. If one restricts to consider just polynomial solutions,  say $f_{r}(\boldsymbol{x}) \equiv p_{r}(\boldsymbol{x})$ for $r=1,\dots, N_p(\mathfrak{g})$, the classical Casimir operators as elements of the universal enveloping algebra $\mathcal{U}(\mathfrak{g})$ are obtained as a result of a symmetrization procedure of the polynomials $p_{r}(\boldsymbol{x})$ after replacing the linear coordinates with the corresponding generators of the Lie algebra, i.e., using the symmetrization map   
\begin{equation}
\begin{split}
\Lambda: {\rm Pol}(\mathfrak{g}^*) & \to \mathcal{U}(\mathfrak{g})\\
p_r(\boldsymbol{x}) & \to \mathsf{P}_r(\boldsymbol{X}):=\Lambda\left(p_r(\boldsymbol{x})\right)
\end{split}\label{symm}
\end{equation}
the explicit action of which on monomials reads:
\begin{equation}
\Lambda\left(x_{i_1} \dots x_{i_n}\right) =\frac{1}{n!} \sum_{\sigma \in \Sigma_n}X_{i_{\sigma(1)}} \dots X_{i_{\sigma(n)}}
\label{symma}
\end{equation}
with $\Sigma_n$ being the permutation group in $\{1, 2, \dots, n\}$. Taking this into account, in the papers \cite{cam07, cam09, cam09m,cam11} this approach has been adapted to the analysis of Lie algebra-subalgebra chains $\mathfrak{g} \supset \mathfrak{g}'$ where, to find the (unsymmetrized) polynomials corresponding to the missing label operators the above system of linear PDEs was restricted to the subalgebra generators. Interestingly enough, in one of these papers the commutativity of the missing label operators was understood in the commutative setting as a consequence of an involution condition obtained in terms of the Lie-Poisson bracket carried by the dual $\mathfrak{g}^*$ of the Lie algebra $\mathfrak{g}$, the latter being given by \cite{per, pois}:
\begin{equation}\label{bere}
\{f,g\}=C_{ij}^k x_k \partial_{x_i} f \partial_{x_j} g  \, ,
\end{equation}
where $f,g \in \textsf{Pol}(\mathfrak{g^*})$ and $\partial_{x_j}f=\frac{\partial f(\boldsymbol{x})}{\partial x_{j}}$ 
in this setting. The coadjoint action of $\mathfrak{g}$ on $\textsf{Pol}(\mathfrak{g^*})$ is \cite{IS84}:
\begin{equation*}
\hat{X}_i(f(\boldsymbol{x}))=C_{i j}^k x_k \partial_{x_j} f(\boldsymbol{x})=\{x_i,f(\boldsymbol{x})\} \qquad f(\boldsymbol{x}) \in \textsf{Pol}(\mathfrak{g}^*) \, .
\end{equation*}
That said, more recent papers have been devoted to the construction of polynomial algebras, mostly quadratic, arising from the analysis of specific commutants of algebraic Hamiltonians defined in the enveloping algebra of a given Lie algebra $\mathfrak{g}$, also non-semisimple \cite{cam21,cor21, cam22, cam22acta}. The purpose of this work is to proceed along these lines, based on the analysis of commutants and the construction of polynomials in the enveloping algebra of Lie algebras, in order to determine polynomial algebras arsing from another types of commutants, more specifically, those associated to a certain Lie subalgebra $\mathfrak{g} '$ and the corresponding reduction chain $\mathfrak{g} \supset \mathfrak{g} '$. This allows us to investigate, among others, chains related to nuclear models and missing labels from a different perspective. In fact, considering our previous discussion, our approach will be mainly focused on the commutative side, so we will be mainly dealing with homogeneous polynomials  in $\mathfrak{g}^*$.  In particular, we  primarily construct polynomial Poisson algebras by relying on the Lie-Poisson bracket in $C^\infty(\mathfrak{g}^*)$. The main point is that the analysis of the \textquotedblleft Poisson setting\textquotedblright provides all the information we need to state our results. The noncommutive setting is in fact obtained as a result of the symmetrization procedure sketched above. The corresponding polynomial structures in the enveloping algebra will differ from those obtained at the classical level due to the presence of lower-order correction terms. These terms arise as a consequence of the fact that the symmetrization map is a homomorphism at the level of the Lie algebra $\mathfrak{g}$, but the homomorphism property does not extend to the whole enveloping algebra $\mathcal{U}(\mathfrak{g})$, giving rise to lower-order correction terms \cite{Dix}. Roughly speaking, once a polynomial structure has been identified in terms of polynomials in $\textsf{Pol}(\mathfrak{g}^*)$, the corresponding polynomial structure in terms of polynomials in $\mathcal{U}(\mathfrak{g})$ is obtained looking at those lower-order terms, since the leading term corresponds to what one gets in the commutative setting.

\medskip
The paper is structured as follows: In Section \ref{sec2}, after briefly reviewing some of the basic notions and definitions we need throughout the paper, we present the analytic procedure to compute polynomials in $\textsf{Pol}(\mathfrak{g}^*)$ from a given subalgebra $\mathfrak{g}^{\prime}$ of a Lie algebra $\mathfrak{g}$ in the embedding chain $\mathfrak{g} \supset \mathfrak{g}'$. We then discuss some structural properties of the polynomial Poisson algebras obtained. The noncommutative setting is reviewied focusing on the role of the symmetrization map, where we explain the main differences with the polynomial structures obtained analytically.
In Section \ref{sec3} we discuss three different  examples of physical relevance. We focus in particular on two chains associated to nuclear models, namely the well-known reduction chain $\mathfrak{su}(3) \supset \mathfrak{so}(3)$ associated to the Elliott model and the (extended) chain $\mathfrak{so}(5) \supset \mathfrak{so}(2) \times \mathfrak{u}(1)$ that has been discussed in the context of the Seniority Model \cite{Racahspec, fanorac, Hec, Helm}. In order to present an example that points out that the method is not restricted to embedding chains of semisimple Lie algebras, we discuss the chain $\hat{S}(3) \supset \mathfrak{sl}(2,\mathbb{R}) \times \mathfrak{so}(2)$ based on the thirteen-dimensional centrally-extended Schr\"{o}dinger algebra in $(3+1)$-dimensional space-time \cite{hag72, Nie72, BarXu81, Dob97}. In Section \ref{Sec4} we provide an explicit example in the noncommutative setting. Specifically, we discuss the Elliott chain $\mathfrak{su}(3) \supset \mathfrak{so}(3)$ in order to show how to determine the polynomial algebra in terms of polynomials in the enveloping algebra $\mathcal{U}(\mathfrak{su}(3))$ from the polynomial structure computed analytically. This illustrative example is used to point out the main formal differences in the results which, as mentioned, consist in the appearance of lower-order correction terms.
Finally, in Section \ref{Sec5} we present some concluding remarks and comment on future prospects of the procedure.

 \section{Polynomial algebras from commutants of Lie subalgebras}
 \label{sec2}
 
\subsection{Commutants in enveloping algebras}
\label{subsec2.1}

In this subsection, we briefly review some fundamental fact on enveloping algebras and commutants. For further details, see e.g. \cite{Dix,cam22,cam22acta,Moeg} and references therein.
\noindent Let $ \mathcal{ U}(\mathfrak{g})$ denote the universal enveloping algebra of an 
 $d$-dimensional Lie algebra $\mathfrak{g}$, and $\{X_1,\dots ,X_d\}$ a basis. If $p$ is a positive integer, we define $\mathcal{U}_{(p)}(\mathfrak{g})$ as the linear space generated by all monomials $X_1^{a_1}\dots X_d^{a_d}$ such that the inequality $a_1+a_2+\dots +a_d\leq p$ is satisfied. This allows to define the degree $q$ of an arbitrary element $P\in \mathcal{U}(\mathfrak{g})$ as $q={\rm inf}\left\{k\;|\; P\in \mathcal{U}_{(k)}(\mathfrak{g})\right\}$. The subspaces $\mathcal{U}_{(p)}(\mathfrak{g})$ determine a natural filtration in the enveloping algebra $ \mathcal{ U}(\mathfrak{g})$, i.e., for integers $p,q\geq 0$ the relations  
\begin{equation}
\mathcal{U}_{(0)}(\mathfrak{g})=\mathbb{K},\quad \mathcal{U}_{(p)}(\mathfrak{g})\mathcal{U}_{(q)}(\mathfrak{g})\subset \mathcal{U}_{(p
+q)}(\mathfrak{g}),\quad \mathcal{U}_{(p)}(\mathfrak{g})\subset \mathcal{U}_{(p+k)}(\mathfrak{g}),\; k\geq 1 \label{fil1}
\end{equation}
are satisfied, where the base field is $\mathbb{K}=\mathbb{R,C}$. It can be shown that the subspaces $\mathcal{U}_{(p)}(\mathfrak{g})$ are finite-dimensional representations of the Lie algebra $\mathfrak{g}$, from which a realization of $\mathcal{U}(\mathfrak{g})$ in terms of the representation theory of $\mathfrak{g}$ can be deduced \cite{Dix}.

\medskip
\noindent The connection of the algebraic formalism with the analytical approach is obtained in terms of the 
adjoint action of $\mathfrak{g}$ on the enveloping and the symmetric algebra $S(\mathfrak{g})$, defined by  
\begin{equation}\label{adja}
\begin{array}[c]{rl}
P\in \mathcal{U}(\mathfrak{g}) \mapsto & P.X_i:= \left[X_i,P\right]= X_i P-P X_i\in\mathcal{U}(\mathfrak{g}),\\
P\left(x_1,\dots ,x_d\right)\in S(\mathfrak{g})\mapsto &\displaystyle  \widehat{X}_i(P)=C_{ij}^{k} x_k \frac{\partial P}{\partial x_j}\in S(\mathfrak{g}),\\
\end{array}
\end{equation}
respectively. The differential operators $\widehat{X}_i$ (see equation (\ref{difo1})) are easily identified with the infinitesimal generators of one-parameter subgroups determined by  the generators $X_i$ by means of the coadjoint representation \cite{Cheb}. The symmetric algebra $S\left( \frak{g}\right)$ can further be identified with the polynomial algebra $\textsf{Pol}(\mathfrak{g}^*)=\mathbb{K}\left[ x_{1},\ldots ,x_{d}\right] $, and by means of the Berezin bracket defined by equation (\ref{bere}), the symmetric algebra inherits the structure of a Poisson algebra that contains a subalgebra isomorphic to $\frak{g}$. Further, the symmetrization map (\ref{symm}) gives rise to the canonical linear isomorphism $\Lambda:S(\mathfrak{g})\rightarrow \mathcal{U}(\mathfrak{g})$ that commutes with the adjoint action. From the identity $\mathcal{U}^{(p)}(\mathfrak{g})=\Lambda\left(S^{(p)}(\mathfrak{g})\right)$, with $S^{(p)}(\mathfrak{g})$ denoting the  homogeneous polynomials of degree $p$, the decomposition $\mathcal{U}_{(p)}(\mathfrak{g})=\sum_{k=0}^{p} \mathcal{U}^{(k)}(\mathfrak{g})$ is deduced. This in particular implies the relation 
\begin{equation*}
\left[ P,Q\right]\in \mathcal{U}_{(p+q-1)}(\mathfrak{g}),\quad P\in \mathcal{U}_{(p)}(\mathfrak{g}), Q\in\mathcal{U}_{(q)}(\mathfrak{g}).
\end{equation*}
The centre 
\begin{equation}\label{INVS1}
Z\left(\mathcal{U}(\mathfrak{g})\right) = \left\{ P\in\mathcal{U}(\mathfrak{g})\; |\;\left[\mathfrak{g},P\right]=0\right\}
\end{equation}
consists of invariant polynomials of $\mathfrak{g}$. Using the commutative frame (see (\ref{adja})), it can be easily seen that the centralizer of the symmetric algebra 
\begin{equation*}
Z\left(S\left( \frak{g}\right)\right)=\left\{ P\in S\left( \frak{g}\right) \;|\;\left\{ P,Q\right\} =0,\;Q\in
S\left( \frak{g}\right) \right\} ,
\end{equation*}
is not only linearly but algebraically isomorphic to $Z\left(\mathcal{U}(\mathfrak{g})\right)$, although, in general, this algebraic isomorphism does not coincide with the symmetrization map $\Lambda$ (see e.g. \cite{Cheb,Dix}).

\medskip
\noindent We define the commutant $C_{\mathcal{U}(\mathfrak{g})}(\mathfrak{g}^{\prime})$ of a subalgebra $\mathfrak{g}^{\prime}\subset \mathfrak{g}$ by means of the condition  
\begin{equation}
C_{\mathcal{U}(\mathfrak{g})}(\mathfrak{g}^{\prime})=\left\{ Q\in\mathcal{U}(\mathfrak{g})\; |\; [P,Q]=0,\quad \forall P\in\mathfrak{g}^{\prime}\right\}.\label{comm}
\end{equation}
Determining the commutant directly in the enveloping algebra is generally a cumbersome task \cite{cam21,cam22}, for which reason it is computationally more efficient to consider the equivalent analytical formulation, using the canonical isomorphism $\Lambda$. In this context, the polynomials in the centralizer 
\begin{equation*}
C_{S(\frak{g})}\left( \frak{g}^{\prime}\right) =\left\{ Q\in S\left( 
\frak{g}\right) \;|\;\left\{ P,Q\right\} =0,\;P\in \frak{g}^{\prime}\right\} .
\end{equation*}
are determined as the solutions of the equations in 
system (\ref{eq:casinv}) corresponding to the subalgebra generators
\begin{equation}
\widehat{X}_{i}\left( Q\right) :=\left\{ x_{i},Q\right\} =C_{ij}^{k}x_{k}%
\frac{\partial Q}{\partial x_{j}}=0,\;1\leq i\leq m=\dim \frak{g}^{\prime}.
\label{mlpa}
\end{equation}
The commutant in the enveloping algebra is obtained from the symmetrization map: 
\begin{equation*}
C_{\mathcal{U}(\mathfrak{g})}(\mathfrak{g}^{\prime})=\Lambda\left( C_{S(\frak{g})}\left( \frak{g}^{\prime}\right)\right).
\end{equation*}
This procedure allows to systematically construct a linear basis of $C_{\mathcal{U}(\mathfrak{g})}(\mathfrak{g}^{\prime})$ (see e.g. \cite{RDIY} for details), i.e., a set $\left\{P_1,\dots ,P_r\right\}$ of polynomials such that any element in the commutant admits an expression of the type  
\begin{equation}\label{bas1}
P_1^{a_1}P_2^{a_2}\dots P_s^{a_s},\quad a_i\in\mathbb{N}\cup {0},
\end{equation}
with the $a_i$ fulfilling some algebraic conditions. These polynomials are not required to be algebraically independent, and we speak of the linear dimension of the commutant as $\dim_{L} C_{\mathcal{U}(\mathfrak{g})}(\mathfrak{g}^{\prime})=s$. The relaxation of the algebraic to the linear independency is justified by the fact that commutators of algebraically independent polynomials may not be expressible in the form (\ref{bas1}), but as rational functions of the basis elements \cite{RDIY}.   

\subsection{Commutative setting: polynomial Poisson algebras from polynomials in $\mathsf{Pol}(\mathfrak{g^*})$}
  \label{sec2.1}

Let us consider the reduction chain $\mathfrak{g} \supset \mathfrak{g}'$, where $\mathfrak{g}$ is a $d$-dimensional Lie algebra spanned by the basis generators $\boldsymbol{X}:=\{X_1, \dots, X_d\}$ satisfying the commutation relations:
\begin{equation}
 [X_i,X_j]=C_{ij}^k X_k  \, ,
 \end{equation}
and $\mathfrak{g}^{\prime}$ is an $m$-dimensional subalgebra. Considering the Lie-Poisson bracket (\ref{bere}) on the dual $\mathfrak{g}^*$ of $\mathfrak{g}$, the basic bracket relations for the coordinates $\boldsymbol{x}=\{x_1, . . . , x_d\}$ are
\begin{equation}
\{x_i,x_j\}=C_{ij}^k x_k  \, ,
\end{equation}
while for arbitrary functions $f,g \in C^\infty(\mathfrak{g}^*)$ we have
\begin{equation}
\{f,g\}=C_{ij}^k x_k \partial_{x_i} f \partial_{x_j} g  \, .
\label{LPB}
\end{equation}
Without loss of generality we can suppose that $\{X_1, \dots, X_m\}$ is a basis of $\mathfrak{g}^{\prime}$, while $\{X_{m+1}, \dots, X_d\}$ generates the complementary of  $\mathfrak{g}^{\prime}$ in $\mathfrak{g}$. In the context of the labelling problem \cite{cam11}, functions (not necessarily polynomials) satisfying the system of PDEs (\ref{mlpa}) correspond to labelling operators, i.e., operators that can be used to separate states whenever a representation of $\mathfrak{g}$ is decomposed with respect to the subalgebra $\mathfrak{g}^{\prime}$ (see e.g. \cite{RacahLect}). The system (\ref{mlpa}) possesses exactly $\mathcal{M}_0=\dim \mathfrak{g}-\dim\mathfrak{g}^{\prime}+\ell_0$ independent solutions, where $\ell_0$ denotes the number of $\mathfrak{g}$-invariants that depend solely on variables of the subalgebra $\mathfrak{g}^{\prime}$. As shown in \cite{Pe76}, the identity 
\begin{equation}
\mathcal{M}_0=2n_0+N\left(\mathfrak{g}\right)+N\left(\mathfrak{g}^{\prime}\right)
\end{equation}
holds, where $n_0$ is the number of required (internal) labelling operators. From this we conclude that there are three types of elements that commute with the subalgebra, namely the Casimir operators of $\mathfrak{g}$, the Casimir operators of $\mathfrak{g}^{\prime}$
and $2n$ genuine labelling operators that depend on both generators of $\mathfrak{g}$ and $\mathfrak{g}^{\prime}$. It can be shown (see e.g. \cite{Pe76}) that among these $2n$ operators, at most $n$ commute with each other. It follows from this observation that the solutions of (\ref{mlpa}) generate a functional algebra with respect to the Poisson bracket. Restricting to the case of polynomials, i.e., to the symmetric algebra (see equation (\ref{symm})), the solutions generate a polynomial algebra that is non-Abelian in general. The symmetrized operators thus generate a polynomial algebra in the enveloping algebra.  

\medskip
This justifies that in the following we focus principally on homogenous polynomials of degree $n \geq 1$ and the generic form 
\begin{equation}
p^{(n)}(\boldsymbol{x})=\sum_{a_1+\dots + a_d = n} c_{a_1, \dots, a_d} \,x_1^{a_1} \dots x_d^{a_d}  \in S\left( \frak{g}\right)=\mathsf{Pol}(\mathfrak{g}^*)\, .
\label{polynomials}
\end{equation}
Suppose the Lie subalgebra $\mathfrak{g}'$ is spanned by a subset $\boldsymbol{X}' \subset \boldsymbol{X}$  composed by $m<d$ generators, each of which has a corresponding linear coordinate on $\mathfrak{g^*}$, the whole subset being $\boldsymbol{x}' \subset \boldsymbol{x}$. For embedding chains, elements in the commutant (\ref{comm}) correspond to  polynomials that Poisson commute with the corresponding linear coordinates associated to the generators of the subalgebra $\mathfrak{g}'$ w.r.t. the Lie-Poisson bracket \eqref{LPB}. Assuming the generic expansion \eqref{polynomials}, this requirement reads 
\begin{equation}
\{x_{\alpha}, p^{(n)}(\boldsymbol{x})\} =0\quad 1\leq \alpha\leq m.
\label{dbd}
\end{equation}
These equations can be solved successively, i.e., analyzing systematically the homogeneous solutions of a certain degree, up to some degree $n=N$ after which we only obtain additional linearly dependent elements, namely polynomials that can be all expressed in terms of combinations of the lower-order ones. For degree $n=1$ the condition provides elements of $\mathfrak{g}$ that actually belong to the centralizer of $\mathfrak{g}^{\prime}$ in $\mathfrak{g}$. Supposed that there are $m_1 \geq 1$ of such linearly independent degree-one polynomials $$\boldsymbol{p}^{(1)}:=\{p_1^{(1)}(\boldsymbol{x}), \dots, p_{m_1}^{(1)}(\boldsymbol{x})\},$$ we proceed the construction by considering degree $n=2$ polynomials. Among the quadratic solutions of  \eqref{dbd}, those of the type $p_a^{(1)}(\boldsymbol{x})p_b^{(1)}(\boldsymbol{x})$ with $1\leq a,b\leq m_1$ must be discarded, as they do not provide new independent elements. Let $$\boldsymbol{p}^{(2)}:=\{p_1^{(2)}(\boldsymbol{x}), \dots, p_{m_2}^{(2)}(\boldsymbol{x})\} \, $$ 
denote the (genuinely) quadratic solutions. Repeating the process, discarding at each time the polynomials that are obtained as combinations of lower degree ones, we obtain a set composed by $m_{[N]}:=\text{Card}(\boldsymbol{p}^{[N]})=m_1+m_2+\dots +m_N$ linearly independent polynomials up to a given degree $N$:
\begin{equation}
\boldsymbol{p}^{[N]}:=\bigcup_{i=1}^N \boldsymbol{p}^{(i)}  \, ,
\label{listN}
\end{equation}
where for each index we have 
$\boldsymbol{p}^{(i)}:=\{p_1^{(i)}(\boldsymbol{x}), \dots,p_{m_i}^{(i)}(\boldsymbol{x})\}$, $i=1, \dots, N$. As mentioned before, the elements in the set (\ref{listN}) are generally not algebraically independent, but these dependence relations involve rational functions, and thus are outside the polynomial algebra $\mathsf{Pol}(\mathfrak{g^*})$ (see \cite{RDIY} for further details). 

\medskip
Our objective is to construct the finitely generated polynomial (Poisson) algebras spanned by these $m_{[N]}$ elements, aiming to characterize all the polynomial relations constructed from the Lie-Poisson bracket computed between all polynomials:
\begin{equation}
\{p_{i_1}^{(d_{i_1})}(\boldsymbol{x}), p_{i_2}^{(d_{i_2})} (\boldsymbol{x})\}=P_{i_1 i_2}^{(d_{i_1}+d_{i_2}-1)}(\boldsymbol{x})  \, ,
\label{polrel}
\end{equation}
where $P_{i_1 i_2}^{(d_{i_1}+d_{i_2}-1)}(\boldsymbol{x})$ are polynomials of degree $d_{i_1}+d_{i_2}-1$ in $x_i$ ($i=1, \dots, d$). Within the construction of such algebras, whenever a relation between the generators $\boldsymbol{p}^{(M)}$ ($2 \leq M \leq N$) of the type
\begin{equation}
\sum_{i=1}^{m_M} p_i^{(M)}=P^{[M]}(\boldsymbol{p}^{(1)}, \dots,\boldsymbol{p}^{(M-1)})
\label{pol}
\end{equation}
is found, we will discard one polynomial from the set $\boldsymbol{p}^{(M)}$, and compute all relations \eqref{polrel} with the remaining elements, as the dropped polynomial, say the one for $i=i^*$ in \eqref{pol}, can be expressed in terms of the remaining as
\begin{equation}
p_{i^*}^{(M)}=P^{[M]}(\boldsymbol{p}^{(1)}, \dots,\boldsymbol{p}^{(M-1)})-p_1^{(M)}-\dots-p_{i^*-1}^{(M)}-p_{i^*+1}^{(M)}-\dots-p_{m_M}^{(M)} \, .
\label{eq:polel}
\end{equation}

\medskip
\noindent Throughout the paper we adopt the following notational convention to indicate polynomials of a given degree. Once the representatives for each subset $\boldsymbol{p}^{(i)}$ ($i=1, \dots, N$) have been found, taking into account the condition \eqref{eq:polel} to eliminate unessential polynomials, elements of degree one will be indicated with the uppercase letter $A_i$, elements of degree two with $B_j$ and so on, following the alphabetical order. Moreover, central elements will be denoted with lowercase letters, such as $a_k, b_l, c_m, \dots$ again following alphabetical order to keep track of the degree of the homogeneous polynomials. This allows us to keep track easily on the degree of both the generators and central elements, and to clearly differentiate the former from the latter. For the sake of clarity, let us suppose that after performing the construction above we come out with the following four subsets composed by $m_a$ degree-one (central), $m_b$ degree-two, $m_c$ degree-three and $m_d$ degree-four polynomials:
\begin{equation}
\{a_1, \dots, a_{m_a}\}  \cup \{B_1, \dots, B_{m_b}\} \cup \{C_1, \dots, C_{m_c}\}  \cup \{D_1, \dots, D_{m_d}\} \, .
\label{sets}
\end{equation}
Then, taking into account the degree of the above polynomials, it is clear that besides $\{a_i, \cdot\}=0$, the relations of degree-three we have to look for can only adopt the following form:
\begin{align}
\quad \{B_i,B_j\}=\sum_{k=1}^{m_c} f_{ij}^k C_k+\sum_{l=1}^{m_a}\sum_{m=1}^{m_b} f_{ij}^{lm}a_l B_m +\sum_{k=1}^{m_a} \sum_{l=1}^{m_a} \sum_{m=1}^{m_a} f_{ij}^{klm}a_k a_l a_m \, .
\end{align}
Such relations will be abbreviated with the compact notation 
\begin{align}
\quad \{\boldsymbol{B},\boldsymbol{B}\} \sim \boldsymbol{C}\boldsymbol{+}   \boldsymbol{a}  \boldsymbol{B}\boldsymbol{+} \boldsymbol{a}^3 
\end{align}
with the aim of specifying the combinations of subsets of a given degree involved in the construction. With this compact notation it is clear that if the polynomial algebra closes at this level, the other relations involving the polynomial elements in the set \eqref{sets} must have the form
\begin{align}
\quad \{\boldsymbol{B},\boldsymbol{C}\} &\sim \boldsymbol{D}\boldsymbol{+}   \boldsymbol{B}^2\boldsymbol{+} \boldsymbol{a} \{\boldsymbol{B}, \boldsymbol{B}\} \nonumber \\
\quad \{\boldsymbol{C},\boldsymbol{C}\} &\sim    \boldsymbol{B}  \boldsymbol{C}\boldsymbol{+} \boldsymbol{a} \{\boldsymbol{B}, \boldsymbol{C}\}\nonumber \\
\{\boldsymbol{C},\boldsymbol{D}\} &\sim \boldsymbol{C}^2\boldsymbol{+}\boldsymbol{B}\boldsymbol{D}\boldsymbol{+}\boldsymbol{B}^3\boldsymbol{+} \boldsymbol{a}\{\boldsymbol{C},\boldsymbol{C}\}\nonumber \\
\{\boldsymbol{D},\boldsymbol{D}\}&\sim \boldsymbol{C}\boldsymbol{D}\boldsymbol{+}\boldsymbol{B}^2\boldsymbol{C}\boldsymbol{+}\boldsymbol{a}\{\boldsymbol{C},\boldsymbol{D}\} \, ,
\end{align}
indicating that expansions up to degree seven are required. As an example, opening up the last formal expression we get 
\begin{align}
\{\boldsymbol{D},\boldsymbol{D}\}&\sim \boldsymbol{C}\boldsymbol{D}\boldsymbol{+}\boldsymbol{B}^2\boldsymbol{C}\boldsymbol{+}\boldsymbol{a}\{\boldsymbol{C},\boldsymbol{D}\} \nonumber \\
&\sim \boldsymbol{C}\boldsymbol{D}+\boldsymbol{B}^2 \boldsymbol{C}+\boldsymbol{a}(\boldsymbol{C}^2\boldsymbol{+}\boldsymbol{B}\boldsymbol{D}\boldsymbol{+}\boldsymbol{B}^3\boldsymbol{+} \boldsymbol{a}\{\boldsymbol{C},\boldsymbol{C}\}) \nonumber \\
& \sim \boldsymbol{C}\boldsymbol{D}+\boldsymbol{B}^2 \boldsymbol{C}+\boldsymbol{a}(\boldsymbol{C}^2\boldsymbol{+}\boldsymbol{B}\boldsymbol{D}\boldsymbol{+}\boldsymbol{B}^3\boldsymbol{+} \boldsymbol{a}(  \boldsymbol{B}  \boldsymbol{C}\boldsymbol{+} \boldsymbol{a} \{\boldsymbol{B}, \boldsymbol{C}\})) \nonumber \\
&\sim \boldsymbol{C}\boldsymbol{D}+\boldsymbol{B}^2 \boldsymbol{C}+\boldsymbol{a}(\boldsymbol{C}^2\boldsymbol{+}\boldsymbol{B}\boldsymbol{D}\boldsymbol{+}\boldsymbol{B}^3\boldsymbol{+} \boldsymbol{a}(  \boldsymbol{B}  \boldsymbol{C}\boldsymbol{+} \boldsymbol{a} (\boldsymbol{D}\boldsymbol{+}   \boldsymbol{B}^2\boldsymbol{+} \boldsymbol{a} \{\boldsymbol{B}, \boldsymbol{B}\}))) \nonumber \\
&\sim \boldsymbol{C}\boldsymbol{D}+\boldsymbol{B}^2 \boldsymbol{C}+\boldsymbol{a}(\boldsymbol{C}^2\boldsymbol{+}\boldsymbol{B}\boldsymbol{D}\boldsymbol{+}\boldsymbol{B}^3\boldsymbol{+} \boldsymbol{a}(  \boldsymbol{B}  \boldsymbol{C}\boldsymbol{+} \boldsymbol{a} (\boldsymbol{D}\boldsymbol{+}   \boldsymbol{B}^2\boldsymbol{+} \boldsymbol{a}(\boldsymbol{C}\boldsymbol{+}   \boldsymbol{a}  \boldsymbol{B}\boldsymbol{+} \boldsymbol{a}^3)))) \nonumber \\
&\sim \boldsymbol{C}\boldsymbol{D}+\boldsymbol{B}^2 \boldsymbol{C}+\boldsymbol{a}(\boldsymbol{C}^2\boldsymbol{+}\boldsymbol{B}\boldsymbol{D}\boldsymbol{+}\boldsymbol{B}^3)\boldsymbol{+} \boldsymbol{a}^2  \boldsymbol{B}  \boldsymbol{C}\boldsymbol{+} \boldsymbol{a}^3 (\boldsymbol{D}\boldsymbol{+}   \boldsymbol{B}^2)\boldsymbol{+} \boldsymbol{a}^4\boldsymbol{C}\boldsymbol{+}   \boldsymbol{a}^5  \boldsymbol{B}\boldsymbol{+} \boldsymbol{a}^7.
\label{formlas}
\end{align}
From this we can easily appreciate that, in principle, all combinations of elements appearing in the last line must be taken into account, 
further indicating their precise structure with respect to the expansion (\ref{polynomials}). According to this prescription, taking for example the first term in the expression above, we immediately see that 
\begin{equation}
\boldsymbol{C}\boldsymbol{D} \qquad \to \qquad \sum_{k=1}^{m_c}\sum_{l=1}^{m_d} f_{ij}^{kl} C_k D_l \qquad (i,j=1, \dots, m_d) \, .
\end{equation}
Clearly, some of the coefficients appearing will be zero. Depending on the embedding chain of Lie algebras considered, additional higher-degree polynomial generators or central elements eventually appear. As a final remark, let us observe that inside the entire set $\boldsymbol{p}^{[N]}$ there can exist combinations of the generators at a given degree that turn out to commute with all the generators involved. When this is the case, these central elements can be used in such a way to select a new basis of the polynomial algebra in which the relations assume a specific form already encountered in the literature in relation to superintegrable systems, missing labels and Clebsch-Gordan problems (see for example \cite{das01,cra20, cra21}). In fact, interestingly enough, when this last passage is implemented for the two examples related to nuclear models, i.e. the Elliott and Senority models, the polynomial algebras that are obtained share several similarities with those types of polynomial structures.

\subsection{Noncommutative setting:  polynomial algebras from polynomials in the enveloping algebra $\mathcal{U}(\mathfrak{g})$}
\label{2.1}

The results obtained in the classical (commutative) setting can be mapped to the noncommutative Lie algebraic setting by means of the canonical isomorphism (\ref{symm}). Supposed that a set $\boldsymbol{p}^{[N]}$ of solutions of \eqref{dbd} composed by $m_{[N]}$ linearly independent polynomials that satisfy the relations \eqref{polrel} has been found, their algebraic counterpart in the enveloping algebra $\mathcal{U}(\mathfrak{g})$ is generated by the images $\Lambda( p^{(n)}(\boldsymbol{x})):=\mathsf{P}^{(n)}(\boldsymbol{X})$. In particular, any monomial gives rise through (\ref{symma}) to a homogeneous polynomial in $\mathcal{U}(\mathfrak{g})$, but where the generators are no more commutative. For computational convenience in the following, we introduce the notation  
\begin{equation}
\Lambda\left(x_{i_1} \cdots x_{i_n}\right)=: \textsf{S}_n(X_{i_1}, X_{i_2},\dots, X_{i_n}),
\label{symmap}
\end{equation}
so that the homogeneity and degree of the polynomials are easily seen:
\begin{equation}
\begin{split}
\textsf{S}_1(X_i)= & X_i,\quad \textsf{S}_2(X_i, X_j)=  \frac{1}{2}(X_i X_j+ X_j X_i),\\
\textsf{S}_3(X_i, X_j,X_k)= & \frac{1}{6}(X_i X_j X_k+X_i X_k X_j+X_j X_i X_k+X_j X_k X_i+X_k X_i X_j+X_k X_j X_i),\quad \dots\\
\end{split}
\end{equation}
From these symmetric representatives it is apparent that, whenever an ordered basis in $\mathcal{U}(\mathfrak{g})$ is chosen, for each $k\geq 2$ the polynomial $\textsf{S}_k(X_{i_1}, X_{i_2},\dots, X_{i_k})$ is rewritten as 
\begin{equation}\label{orba}
\textsf{S}_k(X_{i_1}, X_{i_2},\dots, X_{i_k})=X_{i_1} X_{i_2}\cdots X_{i_k}+ \text{L.O.T.}, 
\end{equation}
in terms of the ordered basis, where L.O.T. denote lower-order terms that arise, after reordering terms, from the non-commutativity of the generators $X_i$. Specifically, assuming that $i\leq j\leq k$ and $n\leq 3$, the representatives in the ordered basis are
\begin{equation}
\begin{split}
\textsf{S}_2(X_i, X_j)= & X_iX_j -\frac{1}{2}\left[X_i,X_j\right]\\
\textsf{S}_3(X_i, X_j,X_k)= &  X_i X_j X_k- \frac{1}{2}X_i\left[X_j,X_k\right]-\frac{1}{2}X_j\left[X_i,X_k\right]-\frac{1}{2}\left[X_i,X_j\right]X_k\\
& +\frac{1}{3}\left[X_i,\left[X_j,X_k\right]\right]+ \frac{1}{6} \left[X_k,\left[X_i,X_j\right]\right].\\
\end{split}
\end{equation}
Hence, for the polynomials 
\begin{equation}
\mathbf{P}^{[N]}:=\bigcup_{i=1}^N \mathbf{P}^{(i)}  \, ,
\label{list4}
\end{equation}
in the commutant, where $\mathbf{P}^{(i)}:=\{\textsf{P}_1^{(i)}(\boldsymbol{X}), \dots,\textsf{P}_{m_i}^{(i)}(\boldsymbol{X})\}$, the corresponding commutation relations adopt the following form \cite{cam09m}:
\begin{equation}
[\textsf{P}_{i_1}^{(d_{i_1})}(\boldsymbol{X}), \textsf{P}_{i_2}^{(d_{i_2})} (\boldsymbol{X})]=\textsf{P}_{i_1 i_2}^{(d_{i_1}+d_{i_2}-1)}(\boldsymbol{X}) + \text{L.O.T.} 
\label{polrelq}
\end{equation}
This points out that the lower-order correction terms determine the main difference between the analytical approach in terms of the Poisson brackets and the algebraic setting within the enveloping algebra $\mathcal{U}(\mathfrak{g})$. This also illustrates that, although 
at the Lie algebra level, the symmetrization map (\ref{symm}) represents a Lie algebra homomorphism, this property does not hold in general anymore at the level of the enveloping algebra $\mathcal{U}(\mathfrak{g})$ \cite{Dix}. As a consequence, there will be differences in the expansion coefficients in the analytical and algebraic settings. However, due to the commutativity of the coordinates, the commutative setting is much more convenient from the computational perspective, as usually polynomials of high degree must be determined. Albeit cumbersome, all conclusions can then be transferred to the enveloping algebra by symmetrization. In addition, and this is an important observation in practice, the information that can be extrapolated, such as the leading-order term of the polynomials in $\mathcal{U}(\mathfrak{g})$ and  the order of the polynomial algebra (quadratic, cubic, quartic and so on) are usually sufficient to formulate our conclusions. Last but not least, the Poisson setting provides useful hints on the general form we have to look for the polynomial algebra in the noncommutative setting. 

\subsection{Abelian subalgebras of $C_{\mathcal{U}(\mathfrak{g})}(\mathfrak{g}^{\prime})$ and the labelling problem}

For any  semisimple Lie algebra $\mathfrak{g}$ of rank $\ell$, the structure theory and the associated root system implies that states within an irreducible representation $\Gamma$  are completely specified by a complete set of 
\begin{equation}
i_0=\frac{1}{2}(\dim \frak{g}+\ell).\label{IntL}\footnote{By equation (\ref{difo2}), we have the identify $\mathcal{N}(\frak{g})=\ell$.}
\end{equation}
labels,  where $\ell$ corresponds to the number of independent Casimir operators, i.e., polynomials in the generators that commute with all elements in $\frak{g}$. The eigenvalues of these operators are used to separate irreducible representations (IR's) of $\frak{g}$, implying that $j_0=\frac{1}{2}(\dim \frak{g}-\ell)$ internal labels are necessary to distinguish states. The generators of the Cartan subalgebra $\mathfrak{h}$ distinguish states within $\Gamma$, but provide only $\ell$ operators, a quantity usually not sufficient to separate degenerate states, so that the computation of further $\frac{1}{2}(\dim \frak{g}-3\ell)$ is required to avoid ambiguities in the description \cite{Pe76}. In the context of polynomial algebras, the (commuting) operators used to label representations of $\mathfrak{g}$ correspond to (commuting) elements of the commutant $C_{\mathcal{U}(\mathfrak{g})}(\mathfrak{h})$.

In physical applications, instead of using the root system, it is customary to use some nonabelian subalgebra $\frak{g}^{\prime}$ (of rank $\ell^{\prime}$) to label the basis states of IR's of $\mathfrak{g}$, pointing out some internal symmetry determined by the embedding $\mathfrak{g}^{\prime}\subset\mathfrak{g}$. In this case, the subgroup provides $\rho_0=\frac{1}{2}(\dim
\frak{g}^{\prime}+\ell^{\prime}))-\ell_{0}$ labels, with $\ell_{0}$ denoting the invariants coomon to $\frak{g}$ and $\frak{g}^{\prime}$. In order to determine a sufficient number of labels (see (\ref{IntL})), it is necessary to determine 
\begin{equation}
n_0=\frac{1}{2}\left(
\dim\frak{g}-\ell-\dim\frak{g}^{\prime}-\ell^{\prime})\right)+\ell_{0}
\label{ML}
\end{equation}
additional commuting operators, usually referred to as missing label operators or subgroup
scalars \cite{Racahspec}. As these operators must commute with the subalgebra $\frak{g}^{\prime}$, they correspond to solutions of the system (\ref{mlpa}), and hence belong to elements in the centralizer $C_{S(\frak{g})}\left(\frak{g}^{\prime}\right)$. Taking into account that the $i_0$ labelling operators must moreover commute with each other in order to be simultaneously diagonalizable \cite{cam07}, it follows that they span a (maximal) Abelian subalgebra of $C_{\mathcal{U}(\frak{g})}\left(\frak{g}^{\prime}\right)$.

\smallskip Besides the Casimir operators that determine the representation, the choice of labelling operators is not canonical, with most of the labelling problems studied in the literature providing the labelling operators of lowest possible order. However, the precise knowledge of the algebraic structure of $C_{\mathcal{U}(\frak{g})}\left(\frak{g}^{\prime}\right)$ (or its analytical counterpart $C_{S(\frak{g})}\left(\frak{g}^{\prime}\right)$) allows to establish the most general possible labelling operators, hence allowing the construction of various (nonequivalent) orthonormal bases of states for IR's of $\mathfrak{g}$.  

\medskip The embedding $\mathfrak{g}^{\prime}\subset \mathfrak{g}$ induces a decomposition of IR's $\Gamma$ of $\mathfrak{g}$ as a direct sum of $\mathfrak{g}^{\prime}$-multiplets
\begin{equation}\label{irde}
\Gamma\downarrow \Omega_1^{\mu_1}\oplus \dots \oplus \Omega_{r}^{\mu_r},
\end{equation}
where $\Omega_i$ denotes IR's of $\mathfrak{g}^{\prime}$ and $\mu_i$ their multiplicity. Following (\ref{IntL}), an arbitrary state in $\Gamma$ is characterized by $i_0$ eigenvalues $\lambda_s$ ($1\leq s\leq i_0$). The commutant $C_{\mathcal{U}(\frak{g})}\left(\frak{g}^{\prime}\right)$ provides essentially three different types of labelling operators:
\begin{enumerate}
\item Casimir operators of $\mathfrak{g}$ determining $\Gamma$ (with $\ell$ independent operators)

\item Casimir operators of $\mathfrak{g}^{\prime}$ determining the nonequivalent IR's appearing in (\ref{irde}), with $\ell^{\prime}$ independent operators. 

\item Missing label operators that separate degeneracies whenever $\mu_s>1$ for some $s\in \{1,\dots ,r\}$. At most $n_0$ of such operators commute with each other. 
\end{enumerate}
In addition to these three types, that determine and separate the representations appearing in the decomposition, we have to consider also $\rho_0$internal operators (typically obtained either considering generators of $\mathfrak{g}^{\prime}$ or some additional subalgebra) that separate the individual states within each IR $\Omega_s$. It follows that any state of $\Gamma$ is completely characterized by the four types of eigenvalues described above, allowing to write the state as the eigenvalue sequence  
\begin{equation}\label{state}
\left | \lambda_1,\dots,\lambda_{\ell}; \lambda_1^{\prime},\dots \lambda_{\ell^{\prime}}^{\prime};\xi_1,\dots ,\xi_{n_0};\chi_1,\dots ,\chi_{\rho_0}
 \right>.
\end{equation}
For specific applications, and whenever there is no ambiguity concerning the representation $\Gamma$, it is also conceivable to replace the Casimir operators of $\mathfrak{g}$ by other elements in the commutant, so that different states can have different eigenvalues. From this more general perspective, any choice of $i_0$ commuting polynomials of $\frak{g}^{\prime}$ and $C_{\mathcal{U}(\frak{g})}\left(\frak{g}^{\prime}\right)$ is potentially of use to describe states of $\mathfrak{g}$ with respect to the subgroup $\mathfrak{g}^{\prime}$. This may also be of interest in the context of dynamical symmetries \cite{iacbook}, with the Hamiltonian of the model described in terms of the Casimir operators of the Lie algebras in the chain. 

\medskip

\section{Explicit examples}
\label{sec3}

In the following, the method to construct polynomial algebras will be illustrated by considering three different Lie algebra embeddings $\mathfrak{g}^{\prime} \subset \mathfrak{g}$, two of which have been found relevant in Nuclear Physics. For computational convenience, as mentioned, we choose the Poisson setting, which provides to us all information we need to establish our results. Specifically, the analysis will be focused on the three following cases (the last column indicating the number $n_0$ of missing label operators):

\begin{table}[h!]
	\centering
	\begin{tabular}{c c c c c} 
		$\mathfrak{g}$ &$\mathfrak{g}^{\prime}$ & $\textsf{dim}(\mathfrak{g})$ & $\textsf{dim}(\mathfrak{g}')$  & $n_0$\\ [0.5ex] 
		\hline 
		$\mathfrak{su}(3)$ & $\mathfrak{so}(3)$ & 8 & 3 & 1\\ [1ex] 
		$\mathfrak{so}(5)$ & $\mathfrak{su}(2) \times \mathfrak{u}(1)$  & 10 & 4 &1  \\ [1ex] 
		$\hat{S}(3)$ & $\mathfrak{sl}(2,\mathbb{R}) \times \mathfrak{so}(2)$ & 13 & 4 &2 \\ [1ex] 
			\end{tabular}
	\caption{Algebra-subalgebra chains $\mathfrak{g} \supset \mathfrak{g}'$}
	\label{table:1}
\end{table}
\noindent Moreover, with the aim to present a guiding example useful to point out the main differences with respect to the noncommutative setting, we also provide the corresonding scenario for the case of the Elliott chain  $\mathfrak{su}(3) \supset \mathfrak{so}(3)$. In this context, we explicitly construct the polynomials in the enveloping algebra $\mathcal{U}(\mathfrak{su}(3))$, together with the corresponding polynomial commutator algebra and the associated Casimir invariant.
\vskip 0.5cm

\subsection{The chain $\mathfrak{su}(3) \supset \mathfrak{so}(3)$: Elliott model}  
\label{sec3.1}

The first example we discuss concerns the chain associated to one of the best studied missing label problems,  the so-called Elliott chain $\mathfrak{su}(3) \supset \mathfrak{so}(3)$,  relevant to the study of the Elliott model in Nuclear Physics \cite{ell1, ell2}.  This chain has been investigated from the point of view of missing labels in the recent work \cite{cra21}, where the reader can also find a comprehensive list of older and other pertinent references related to the subject.  In this chain, the $\mathfrak{so}(3)$ subalgebra is spanned by the three orbital angular momentum operators $\boldsymbol{L}=(L_1, L_2, L_3)$, while the remaining five generators are the components of a rank-two symmetric traceless tensor:
\begin{equation}
\boldsymbol{T}=\begin{pmatrix}
T_{11} & T_{12}  & T_{13}  \\ 
T_{12} & T_{22}  & T_{23}  \\  
T_{13}  & T_{23} & -T_{11} -T_{22}    
\end{pmatrix} \, .
\end{equation}
The commutation relations associated to this basis  can be easily derived starting from the standard $\mathfrak{su}(3)$-basis 
\begin{equation}
[E_{ij}, E_{kl}]=\delta_{jk}E_{il}-\delta_{il}E_{kj}\quad (i,j,k,l=1,2,3)\qquad  \cup \qquad E_{11}+E_{22}+E_{33}=0 
\end{equation}
through the following change of basis \cite{RacahLect, cra21}:
\begin{equation}\label{subTij}
\begin{split}
L_1&=\frac{E_{12}+E_{21}-E_{23}-E_{32}}{\sqrt{2}} \, ,\qquad L_2=- {\rm i} \frac{E_{12}-E_{21}+E_{32}-E_{23}}{\sqrt{2}} \, ,\qquad L_3=E_{11}-E_{33} \\ 
T_{11}&=E_{11}+E_{13}+E_{31}+E_{33} \, , \qquad T_{12}=- {\rm i} (E_{13}-E_{31} ) \, ,  \qquad T_{13}=- \frac{E_{12}+E_{21}+E_{23}+E_{32}}{\sqrt{2}}\\
T_{22}&=E_{11}-E_{13}-E_{31}+E_{33} \, , \qquad T_{23}=\frac{E_{12}-E_{21}+E_{23}-E_{32}}{\sqrt{2}}, \qquad {\rm i}=\sqrt{-1},
\end{split}
\end{equation}
leading to the commutation relations:
\begin{equation}\label{rel}
\begin{split}
[ L_{i},L_{j}]&= {\rm i}\epsilon_{ijk}L_k,\quad   [ L_{i},T_{jk}]= {\rm i}\epsilon_{ijl} T_{kl} +{\rm i}  \epsilon_{ikl} T_{jl},\\ 
[ T_{ij}, T_{kl}]&= {\rm i} (\delta_{ki} \epsilon_{jlm} + \delta_{li} \epsilon_{jkm}+ \delta_{kj} \epsilon_{ilm} + \delta_{lj} \epsilon_{ikm} ) L_m.
\end{split}
\end{equation}
The explicit commutator table in the $(\boldsymbol{L},\boldsymbol{T})$ basis is given by:
\begin{center}
{\small
\begin{tabular}{c|c|c|c|c|c|c|c|c}
${[\cdot,\cdot]}$ & ${L_1}$ & ${L_2}$& ${L_3}$ & ${T_{11}}$ & ${T_{12}}$ & ${T_{13}}$ & ${T_{22}}$ & ${T_{23}}$ \\	\hline
${L_1}$  &$0$ & ${\rm i} L_3$ & $-{\rm i} L_2$ & $0$ & ${\rm i} T_{13}$ & $-{\rm i} T_{12}$ & $2 {\rm i} T_{23}$ & $-{\rm i} T_{11}-2 {\rm i} T_{22}$ \\	\hline
${L_2}$ & $-{\rm i} L_3$ & $0$ & ${\rm i} L_1$ & $-2 {\rm i} T_{13}$ & $-{\rm i} T_{23}$ & $2 {\rm i} T_{11}+{\rm i} T_{22}$ & $0$ & ${\rm i} T_{12}$ \\	\hline
${L_3}$  & ${\rm i} L_2$ & $-{\rm i} L_1$ & $0$ & $2 {\rm i} T_{12}$ & $i T_{22}-{\rm i} T_{11}$ & ${\rm i} T_{23}$ & $-2 {\rm i} T_{12}$ & $-{\rm i} T_{13}$ \\	\hline
${T_{11}}$  & $0$ & $2 {\rm i} T_{13}$ & $-2 {\rm i} T_{12}$ & $0$ & $2 {\rm i} L_3$ & $-2 {\rm i} L_2$ & $0$ & $0$ \\	\hline
${T_{12}}$ & $-{\rm i} T_{13}$ & ${\rm i} T_{23}$ & ${\rm i} T_{11}-{\rm i} T_{22}$ & $-2 {\rm i} L_3$ & $0$ & ${\rm i} L_1$ & $2 {\rm i} L_3$ & $-{\rm i} L_2$ \\	\hline
${T_{13}}$ & ${\rm i} T_{12}$ & $-2 {\rm i} T_{11}-{\rm i} T_{22}$ & $-{\rm i} T_{23}$ & $2 {\rm i} L_2$ & $-{\rm i} L_1$ & $0$ & $0$ & ${\rm i} L_3$ \\	\hline
${T_{22}}$ & $-2 {\rm i} T_{23}$ & $0$ & $2 {\rm i} T_{12}$ & $0$ & $-2 {\rm i} L_3$ & $0$ & $0$ & $2 {\rm i} L_1$ \\	\hline
${T_{23}}$ & ${\rm i} T_{11}+2 {\rm i} T_{22}$ & $-{\rm i} T_{12}$ & ${\rm i} T_{13}$ & $0$ & ${\rm i} L_2$ & $-{\rm i} L_3$ & $-2 {\rm i} L_1$ & $0$ \\	\hline
\end{tabular} }
\end{center}
\noindent From these relations, we define the Lie-Poisson bracket \eqref{LPB} in terms of the linear coordinates: 
\begin{equation*} \boldsymbol{x}:=\{x_1,x_2,x_3,x_4,x_5,x_6,x_7,x_8\} \equiv \{\ell_1, \ell_2, \ell_3, t_{11},t_{12},t_{13},t_{22},t_{23}\} \, .
\end{equation*}
\noindent In the Poisson setting, the commutators \eqref{rel} translate to the following ones:
\begin{align}
\{\ell_{i},\ell_{j}\}&= {\rm i} \epsilon_{ijk}\ell_k \, , \hskip 1cm
\{\ell_{i},t_{jk}\}= {\rm i} \epsilon_{ijl} t_{kl} + {\rm i} \epsilon_{ikl} t_{jl} \nonumber \\
\{t_{ij}, t_{kl}\}&= {\rm i} (\delta_{ki} \epsilon_{jlm} + \delta_{li} \epsilon_{jkm}+ \delta_{kj} \epsilon_{ilm} + \delta_{lj} \epsilon_{ikm} ) \ell_m \label{ee6} \, .
\end{align}

\noindent We now proceed with our systematic procedure to find the classical (unsymmetrized) elements defining the commutant of the rotation subalgebra corresponding to the coordinates $\boldsymbol{x}':=\{x_1, x_2, x_3\}\equiv\{\ell_1, \ell_2, \ell_3\}$.
Given the set of elements $\boldsymbol{x}=\{\ell_{1}, \ell_{2},\ell_{3},t_{11}, t_{12}, t_{13}, t_{22}, t_{23}\}$, the expansion (\ref{polynomials}) at order $n$ is given by 
\begin{equation}
p^{(n)}(\boldsymbol{x})=\sum_{a_1+\dots + a_8= n} c_{a_1, \dots, a_8} \,x_1^{a_1} \dots x_8^{a_8} 
\label{eq:exp1}
\end{equation}
with the commutant constraint
\begin{equation}
\{\ell_k, p^{(n)}(\boldsymbol{x})\}=0,\quad k=1,2,3 .
\label{pdes}
\end{equation}

\noindent As the embedding $\mathfrak{so}(3)\subset \mathfrak{su}(3)$ is irreducible, it follows at once that the system has no nontrivial solutions for $n=1$, hence there do not exist degree-one elements that Poisson commute with $\{\ell_1, \ell_2, \ell_3\}$ simultaneously. Now the system (\ref{mlpa}) admits five (functionally) independent solutions, implying that the polynomial algebra associated to this reduction chain must have at least five generators. The analysis of the quadratic expansion involves the following $36$ (homogeneous) terms \begin{equation}
\begin{split}
 \{&\ell_1^2, \ell_1 \ell_2, \ell_1 \ell_3, \ell_1 t_{11}, \ell_1 t_{12}, \ell_1 t_{13}, \ell_1 t_{22}, \ell_1 t_{23}, \ell_2^2, \ell_2 \ell_3, \ell_2 t_{11}, \ell_2 t_{12}, \ell_2 t_{13}, \ell_2 t_{22}, \ell_2 t_{23}, \ell_3^2, \ell_3 t_{11}, \ell_3 t_{12}, \ell_3 t_{13},  \\
& \ell_3 t_{22}, \ell_3 t_{23}, t_{11}^2, t_{11} t_{12}, t_{11} t_{13}, t_{11} t_{22}, t_{11} t_{23}, t_{12}^2, t_{12} t_{13}, t_{12} t_{22}, t_{12} t_{23}, t_{13}^2, t_{13} t_{22}, t_{13} t_{23}, t_{22}^2, t_{22} 
t_{23}, t_{23}^2\} \, .
\end{split}
\end{equation}
The constraint (\ref{pdes}) leads to the two homogeneous polynomials
\begin{equation}
\begin{split}
p_1^{(2)}(\boldsymbol{x})&:=\ell_1^2 + \ell_2^2 + \ell_3^2, \\
p_2^{(2)}(\boldsymbol{x})&:=t_{11}^2 + t_{12}^2 + t_{13}^2 + t_{11} t_{22} + t_{22}^2 + t_{23}^2,
\end{split}
\end{equation}
where it is easily seen that $p_1^{(2)}$ corresponds to the Casimir operator of $\mathfrak{so}(3)$ and $p_1^{(2)}+2 p_2^{(2)}$ corresponds to the quadratic Casimir operator of $\mathfrak{su}(3)$. The third-order expansion (\ref{eq:exp1})  involves $120$ homogeneous cubic terms. Solving the corresponding system (\ref{pdes}) for the undetermined coefficients, we get the  following two cubic polynomials as solutions:
\begin{equation}
\begin{split}
p_1^{(3)}&:= \ell_3^2 (t_{11} + t_{22}) -\ell_1^2 t_{11} - 2 \ell_1 (\ell_2 t_{12} + \ell_3 t_{13}) - \ell_2^2 t_{22} - 
2 \ell_2 \ell_3 t_{23} \\
p_2^{(3)}&:= t_{12} t_{13} t_{23}+\frac{1}{2} \left( t_{11} (t_{12}^2 - t_{22}^2 - t_{23}^2)-t_{22}(t_{11}^2  - t_{12}^2  + t_{13}^2)\right) \, . 
\end{split}
\end{equation}
It is straightforward to verify that $p_1^{(3)}+2 p_2^{(3)}$ coincides with the cubic Casimir operator of $\mathfrak{su}(3)$. 
Proceeding in raising the degree of the expansion, for degree four polynomials four solutions of the system (\ref{pdes}) are found, from which two are discarded, as they arise as the square of $p_1^{(2)}$ and $p_2^{(2)}$, respectively. The remaining solutions are given by:
\begin{equation}
\begin{split}
p_1^{(4)}= &\; \ell_1^2 (t_{11}^2 + t_{12}^2 + t_{13}^2) + 2 \ell_2 \ell_3 (t_{12} t_{13} - t_{11} t_{23}) + 
2 \ell_1 \left(\ell_2 (t_{12} (t_{11} + t_{22}) +  t_{13} t_{23}) +\ell_3(t_{12} t_{23}- t_{13} t_{22} )\right)\\
& + \ell_2^2 (t_{12}^2 + t_{22}^2 + t_{23}^2) + \ell_3^2 (t_{13}^2 + (t_{11} + t_{22})^2 + t_{23}^2), \\ 
p_2^{(4)}= &\; \ell_3^2 (t_{12}^2 - t_{11} t_{22}) + \ell_2^2 (t_{13}^2 + t_{11} (t_{11} + t_{22})) + 2 \ell_1 \ell_3 (t_{13} t_{22} - t_{12} t_{23}) + \ell_1^2 (t_{22} (t_{11} + t_{22}) + t_{23}^2) \\
& - 2 \ell_2 (\ell_1 t_{12} (t_{11} + t_{22}) + \ell_1 t_{13} t_{23} + \ell_3 (t_{12} t_{13} - t_{11} t_{23})) \, .
 \end{split}
\end{equation}
In terms of the labelling problem, either $p_1^{(3)}$ or $p_2^{(3)}$ could be chosen as labelling operators in order three, or alternatively $p_1^{(4)}$ or $p_2^{(4)}$ in order four (see e.g. \cite{mpsw}). By the properties of labelling operators, these cubic and quartic polynomials cannot commute with each other. 

\smallskip
\noindent Solutions of degree five turn out to be all expressible as functions of the lower-order polynomials, so that no new polynomials  arise. For degree six, only one among the ten solutions of the system (\ref{pdes}) results to be independent from the lower-order ones. In particular, this polynomial can be expressed in terms of the Poisson bracket as 
\begin{equation}
p_1^{(6)}:=\frac{{\rm i}}{4} \{p_1^{(3)}, p_1^{(4)}\}=-\frac{{\rm i}}{4} \{p_1^{(3)}, p_2^{(4)}\}= \frac{{\rm i}}{2} \{p_2^{(3)}, p_2^{(4)}\}=-\frac{{\rm i}}{2} \{p_2^{(3)}, p_1^{(4)}\} \, .
\label{eq:polynomial6}
\end{equation}
We observe that $p_1^{(6)}$ is a natural consequence of the non commutativity of the labelling operators, as commented above.

\noindent By direct computation, it can be shown that beyond degree six, no new solutions of the system (\ref{pdes}) that are independent from the polynomials enumerated above arise, indicating that the construction terminates at degree $N=6$. In fact, the polynomial algebra (that is cubic as we will show below) closes in terms of the following seven elements:
\begin{equation}
\{b_1, b_2, C_1, C_2, D_1, D_2, F_1\}:=\{p_1^{(2)}, p_2^{(2)},p_1^{(3)},p_2^{(3)},p_1^{(4)},p_2^{(4)},p_1^{(6)}\} \, ,
\label{polsu3}
\end{equation}
where we have indicated the polynomials using the notation  explained in Section \ref{sec2}. In particular, we notice that the degree-two polynomials turn out to Poisson commute with all the other generators in the set above, i.e. they play the role of central elements for the polynomial algebra: $$\{b_1, \cdot\}=\{b_2, \cdot\}=0 \, .$$

\medskip
\noindent The preceding results can be expressed uniformly using the matrix notation 
	\begin{equation}
	\boldsymbol{\ell}:=\begin{pmatrix}
	\ell_{1}   \\ 
	\ell_{2}  \\  
	\ell_{3}    
	\end{pmatrix} \, ,
	\qquad
	\boldsymbol{t}:=\begin{pmatrix}
	t_{11} & t_{12}  & t_{13}  \\ 
	t_{12} & t_{22}  & t_{23}  \\  
	t_{13}  & t_{23} & -t_{11} -t_{22}    
	\end{pmatrix} \, ,
	\end{equation}
allowing us to characterize the polynomial elements \eqref{polsu3} in terms of trace operators and inner products:  
\begin{equation}
\begin{split}
b_1&=\|\boldsymbol{\ell}\|^2,\quad b_2=\frac{1}{2} \text{Tr}(\boldsymbol{t}^2),\quad  C_1=-(\boldsymbol{t}\,\boldsymbol{\ell})\cdot \boldsymbol{\ell}^{\footnotesize{T}},\quad C_2=\frac{1}{6} \text{Tr}(\boldsymbol{t}^3),\\
D_1 &=\|\boldsymbol{t}\,\boldsymbol{\ell}\|^2 \quad D_2=\frac{1}{2}\|\boldsymbol{\ell}\|^2 \text{Tr} (\boldsymbol{t}^2)-\|\boldsymbol{t}\,\boldsymbol{\ell}\|^2.
\end{split}
\end{equation}	 

\medskip
\noindent For later convenience, we rescale the two polynomials $C_1$ and $F_1$  as $\bar{C}_1:=\frac{1}{2} C_1$ and $\bar{F}_1:=2 F_1$. The brackets of the polynomial algebra are thus given by:
\begin{equation}
\begin{split}
 \{\bar{C}_1, C_2\}&=0,\quad \{\bar{C}_1,D_1\}= - {\rm i} \bar{F}_1,\quad \{\bar{C}_1,D_2\}= + {\rm i} \bar{F}_1,\quad \{D_1, D_2\}=0\\
  \{\bar{C}_1,F_1\}&=  {\rm i} \left( 6 D_1^2+24 b_1\bar{C}_1 C_2-4b_2(2\bar{C}_1^2+ b_1  D_1) \right),\quad \{C_2,D_1\}=+{\rm i} \bar{F}_1,\\
  \{C_2,D_2\}&=-{\rm i} \bar{F}_1,\quad \{C_2, \bar{F}_1\}=-{\rm i}  \left( 6 D_1^2+24 b_1\bar{C}_1 C_2-4b_2(2\bar{C}_1^2+ b_1  D_1) \right),\\
\{D_1,\bar{F}_1\}&=-8{\rm i} \left(4 \bar{C}_1^2(\bar{C}_1-3 C_2)+3 b_1 (\bar{C}_1 F_1 + C_2 D_1)+2 b_2 \bar{C}_1 D_2  - b_1^2 b_2  (\bar{C}_1+2 C_2)  -2 b_1^3 C_2 \right),\\
\{D_2,\bar{F}_1\}&=+8{\rm i} \left(4 \bar{C}_1^2(\bar{C}_1-3 C_2)+3 b_1 (\bar{C}_1 F_1 + C_2 D_1)+2 b_2 \bar{C}_1 D_2  - b_1^2 b_2  (\bar{C}_1+2 C_2)  -2 b_1^3 C_2 \right)\, .  
\end{split}
\end{equation}
From these relations, it can be easily inferred that there exists an algebraic dependency between the fourth-order polynomials: 
\begin{equation}
D_1+D_2-b_1b_2=0,
\label{funrel}
\end{equation}
showing that either $D_1$ or $D_2$ is superfluous and can thus be skipped. In addition, the polynomial $\bar{C}_1+C_2$ commutes with all generators in (\ref{polsu3}) 
\begin{equation}
\{\bar{C}_1+C_2 , \cdot \}=0 \, ,
\label{eq:commp}
\end{equation}
meaning that it corresponds to a central element of the polynomial algebra. This suggests to consider a new basis formed by the following six elements (the seventh having been discarded due to (\ref{funrel})):
\begin{equation}\label{eq:basis}
\begin{split}
c_1&:=b_1 \, , \quad c_2:=b_2\, , \quad c_3:= \bar{C}_1+ C_2 \, , \quad  \mathcal{A}:=\bar{C}_1-C_2 \, ,\quad  \mathcal{B}:=D_1 \, ,\quad  \mathcal{C}:=-2 {\rm i} \bar{F}_1  \, .
\end{split}
\end{equation}
To avoid any confusion, we explicitly list the degree of each polynomial involved in \eqref{eq:basis}:
$$
\textsf{deg}(c_1)=2  \qquad \textsf{deg}(c_2)=2   \qquad \textsf{deg}(c_3)=3  \qquad \textsf{deg}(\mathcal{A})=3  \qquad \textsf{deg}(\mathcal{B})=4 \qquad \textsf{deg}(\mathcal{C})=6 \, . 
$$
In terms of these generators, the three-generator polynomial cubic Poisson algebra below is obtained:
\begin{equation}\label{eqc1}
\begin{split}
\{\mathcal{A},\mathcal{B}\}&=\mathcal{C}  \\
\{\mathcal{A},\mathcal{C}\}&=\alpha \mathcal{A}^2+\beta \mathcal{B}^2 +\delta \mathcal{A}+\epsilon \mathcal{B} +\zeta  \\
\{\mathcal{B},\mathcal{C}\}&=\lambda \mathcal{A}^3+\mu \mathcal{A}^2+2\xi \mathcal{A} \mathcal{B}+\rho \mathcal{A}+\sigma \mathcal{B} +\chi  
\end{split}
\end{equation}
with the coefficients 
\begin{equation}\label{clst1}
\begin{split}
\alpha & =-\xi=-8(3 c_1+c_2) \, , \quad \beta=24 \, , \quad \delta=-\sigma=-16 c_2 c_3 \, , \quad \epsilon=-16 c_1 c_2 \, , \quad \zeta=8 c_3^2(3 c_1-c_2),\\
 \lambda & =-32,\quad \mu=-48 c_3,\quad \rho=-16 c_1(c_1+c_2)^2,\quad \chi= 16 c_3 (c_1(c_1^2 -  c_2^2) + c_3^2) .
\end{split}
\end{equation}
It is routinely verified that the Jacobi identity 
\begin{equation}
\{\mathcal{A},\{\mathcal{B}, \mathcal{C}\}\}+\{\mathcal{C},\{\mathcal{A}, \mathcal{B}\}\}+\{\mathcal{B},\{\mathcal{C}, \mathcal{A}\}\}=0
\label{jac}
\end{equation}
is trivially satisfied. The cubic algebra (\ref{eqc1}) admits a Casimir invariant of the fourth order, explicitly given by 
\begin{equation}\label{ccas}
\begin{split}
\mathcal{K} &=\mathcal{C}^2+\frac{\lambda }{2}\mathcal{A}^4+\frac{2}{3} \mu \mathcal{A}^3 -\frac{2}{3}\beta \mathcal{B}^3-2 \alpha \mathcal{A}^2 \mathcal{B}+\rho \mathcal{A}^2-\epsilon \mathcal{B}^2
-2\delta \mathcal{A} \mathcal{B}+ 2 \chi \mathcal{A}- 2\zeta \mathcal{B},  
\end{split}
\end{equation}
 which in terms of the central elements can be reformulated as
\begin{equation}
K=16 c_3^2(c_1 (c_1 - c_2)^2  -  c_3^2) \, .
\label{eq:central}
\end{equation}
In this context, it is worthy to be mentioned that the Casimir invariant can also be expressed as
\begin{equation}
K=K(\mathcal{A},\mathcal{B},\mathcal{C})=\mathcal{C}^2-2h(\mathcal{A},\mathcal{B}) \, ,
\end{equation} 
with the quartic polynomial in the variables $\mathcal{A},\mathcal{B}$ given by
\begin{equation}
h(\mathcal{A},\mathcal{B})=-\frac{\lambda }{4}\mathcal{A}^4-\frac{1}{3} \mu \mathcal{A}^3 +\frac{1}{3}\beta \mathcal{B}^3+ \alpha \mathcal{A}^2 \mathcal{B}-\frac{1}{2}\rho \mathcal{A}^2+\frac{1}{2}\epsilon \mathcal{B}^2
+\delta \mathcal{A} \mathcal{B}- \chi \mathcal{A}+\zeta \mathcal{B} \, .
\end{equation}
This polynomial plays the role of a ``generating function" of the cubic algebra \eqref{eqc1}, as is inferred from the relations 
\begin{equation}
\{\mathcal{A},\mathcal{B}\}=\mathcal{C},\quad \{\mathcal{A},\mathcal{C}\}=\frac{\partial h(\mathcal{A},\mathcal{B})}{\partial \mathcal{B}},\quad \{\mathcal{B},\mathcal{C}\}=-\frac{\partial h(\mathcal{A},\mathcal{B})}{\partial \mathcal{A}} \, .
\end{equation}
The property of being described in terms of an appropriate generating function has been widely used in the frame of polynomial Poisson algebras associated to superintegrable systems (see e.g. \cite{das01}), which suggests that polynomial algebras related to reduction chains of Lie algebras could constitute an alternative approach to the superintegrability problem \cite{RDIY}. 

\medskip
In the context of the internal labelling problem, for the chain $\mathfrak{so}(3)\subset\mathfrak{su}(3)$ we need $i=3$ internal labels when describing (irreducible) $\mathfrak{su}(3)$-multiplets in terms of $\mathfrak{so}(3)$, according to formula (\ref{ML}), with any representation $\Gamma$ of $\mathfrak{su}(3)$ being characterized by five labels. Suitable choices for orthonormal bases can be extracted from the polynomial algebra (\ref{eqc1}). 
Clearly the Casimir operators $b_1+2b_2$, $\overline{C}_1+C_2$ of $\mathfrak{su}(3)$ have a constant value for all states of $\Gamma$, while the Casimir operator $b_1$ of $\mathfrak{so}(3)$ distinguishes the representations of the latter appearing in the decomposition of $\Gamma$. As internal label we can consider any of the generators of $\mathfrak{so}(3)$,\footnote{As these correspond to rotations, they can be diagonalized over the complex field} with a suitable missing labelling operator $\mathcal{O}$ to separate degeneracies to be chosen as
\begin{equation*}
\mathcal{O}= \Phi\left( \mathcal{A},\mathcal{B},\mathcal{C}\right)
\end{equation*}
with $\Phi$ being an arbitrary function of its arguments. These five operators are independent and commute with each other, and thus provide an orthogonal basis for the states in $\Gamma$.

\subsection{The chain $\mathfrak{so}(5) \supset \mathfrak{su}(2) \times \mathfrak{u}(1)$}
\label{sec3.2}

\noindent The reduction $\mathfrak{so}(5) \supset \mathfrak{su}(2)$ also constitutes a chain that has found various applications in nuclear physics, such as e.g. the surface quadrupole vibrations or the nuclear seniority model, in the context of pairing forces between particles in the same nuclear shell, from which a connection between the Wigner coefficients involving the standard representation of $\mathfrak{so}(5)$ and the fractional parentage coefficients of spin-$2$ systems in the Seniority scheme emerge \cite{Racahspec, fanorac, Hec, Helm}.\footnote{It should however be observed that the embedding is different in each case.} Bases of states for the latter model have been studied in \cite{Sh70}, suggesting an extended labelling problem to the reduction chain $\mathfrak{so}(5) \supset \mathfrak{su}(2) \times \mathfrak{u}(1)$, in order to use the $\mathfrak{u}(1)$ generator as an additional label to separate degenerate states. For this reduction chain, the system (\ref{mlpa}) possesses six (functionally) independent solutions, hence, a polynomial algebra associated to this reduction chain must possess at least six generators. Four of these solutions correspond to the Casimir operators of $\mathfrak{so}(5)$ and $\mathfrak{su}(2) \times \mathfrak{u}(1)$, respectively, so that two additional functions are necessary. As these correspond to genuine labelling operators and only one of such operators is required, they cannot commute with respect to the Poisson bracket (\ref{LPB}). 

\medskip
In this paragraph we reconsider the chain $\mathfrak{so}(5) \supset \mathfrak{su}(2) \times \mathfrak{u}(1)$ from the perspective of polynomial algebras. 
The ten-dimensional Lie algebra $\mathfrak{so}(5)$ is given in terms of the basis $\{S_-, T_-, U_-, V_-, S_+, T_+, U_+,V_+,U_3, V_3\}$ and the $\mathfrak{su}(2) \times \mathfrak{u} (1)$ subalgebra is spanned by the generators $\{U_-, U_+, U_3, V_3\}$, with commutation relations 
\begin{center}
\begin{tabular}{c|cccccccccc}
	${[\cdot\; ,\; \cdot]}$ & ${S_-}$ & ${T_-}$  & ${U_-}$ & ${V_-}$ & ${S_+}$ & ${T_+}$ & ${U_+}$ & ${V_+}$ & ${U_3}$ & ${V_3}$\\ \hline
${S_-}$ & $0$ & $0$ & $0$ & $0$ & $-U_3-V_3$ & $0$ & $-V_-$ & $U_-$ & $S_-$ & $S_-$ \\   
${T_-}$ & $0$ & $0$ & $0$ & $U_-$ & $0$ & $V_3-U_3$ & $-V_+$ & $0$ & $T_-$ & $-T_-$ \\   
${U_-}$ & $0$ & $0$ & $0$ & $2S_-$ & $-V_+$ & $-V_-$ & $-2 U_3$ & $2 T_-$ & $U_-$ & $0$ \\  
${V_-}$ & $0$ & $-U_-$ & $-2 S_-$ & $0$ & $U_+$ & $0$ & $2 T_+$ & $-2 V_3$ & $0$ & $V_-$ \\  
${S_+}$ & $U_3+V_3$ & $0$ & $V_+$ & $-U_+$ & $0$ & $0$ & $0$ & $0$ & $-S_+$ & $-S_+$ \\  
${T_+}$ & $0$ & $U_3-V_3$ & $V_-$ & $0$ & $0$ & $0$ & $0$ & $-U_+$ & $-T_+$ & $T_+$ \\  
${U_+}$ & $V_-$ & $V_+$ & $2 U_3$ & $-2T_+$ & $0$ & $0$ & $0$ & $-2 S_+$ & $-U_+$ & $0$ \\  
${V_+}$ & $-U_-$ & $0$ & $-2 T_-$ & $2 V_3$ & $0$ & $U_+$ & $2 S_+$ & $0$ & $0$ & $-V_+$ \\  
${U_3}$ & $-S_-$ & $-T_-$ & $-U_-$ & $0$ & $S_+$ & $T_+$ & $U_+$ & $0$ & $0$ & $0$ \\  
${V_3}$ & $-S_-$ & $T_-$ & $0$ & $-V_-$ & $S_+$ & $-T_+$ & $0$ & $V_+$ & $0$ & $0$ \\ \hline
\end{tabular}
\end{center}

\noindent With respect to the commutative coordinates: $$\boldsymbol{x}:=\{x_1,x_2,x_3,x_4,x_5,x_6,x_7,x_8, x_9,x_{10}\} \equiv\{s_-,t_-,u_-,v_-,s_+,t_+,u_+,v_+,u_3, v_3\},$$ the subalgebra is spanned by: $$\boldsymbol{x}':=\{x_3, x_7, x_9,x_{10}\}\equiv\{u_-,u_+, u_3, v_3\}, $$
and the commutant is obtained from the system 
\begin{equation}
\{u_\pm, p^{(n)}(\boldsymbol{x})\} =0, \quad \{u_3,\, p^{(n)}(\boldsymbol{x})\},\quad 
\{v_3, \hskip 0.1cm p^{(n)}(\boldsymbol{x})\}=0. 
\label{commso5}
\end{equation}
 Following the implementation of the procedure gives rise to the following nine polynomials up to degree six:
\begin{equation}\label{BA1}
\begin{split}
 a_1&=v_3\\
b_1&=u_3^2+u_- u_+\\
b_2&=2(s_- s_++t_- t_+)+v_- v_+\\
C_1&=2(s_- s_+-t_- t_+)u_3+(t_- u_+-s_+ u_-)v_-+(t_+ u_--s_- u_+)v_+\\
D_1&=8 (s_-^2 s_+^2+t_-^2 t_+^2+(s_-s_++t_-t_+)v_-v_+)+v_-^2v_+^2-4(s_+t_-v_-^2+s_-t_+ v_+^2)\\
D_2&=(4 s_-t_++v_-^2)(4 s_+t_-+v_+^2)\\
D_3&= 2(v_- v_+ - 2 (s_- s_+ + t_- t_+))u_3^2 + 
4 (s_+ u_- v_- + t_- u_+ v_- + t_+ u_- v_+ + s_- u_+ v_+) u_3   \\
& \hskip 0.35cm+ 4 s_+ t_+ u_-^2 + 4 s_- t_- u_+^2 -2 u_- v_- u_+ v_+ \\
F_1&=\bigl((v_-^2 - 4 s_- t_+) u_3^2 + 4 ( u_- t_+  + s_- u_+ ) v_- u_3 + 
2 (t_+^2 u_-^2 + s_-^2 u_+^2) - u_- u_+ v_-^2\bigl) (4 s_+ t_- + v_+^2)\\
F_2&=\bigl((v_+^2 - 4 s_+ t_-) u_3^2 + 4 ( u_- s_+ + t_- u_+ ) v_+ u_3 + 2 (s_+^2 u_-^2 + t_-^2 u_+^2) -  u_- u_+ v_+^2\bigl) (4 s_- t_+ + v_-^2) \,  .
\end{split}
\end{equation}
Higher-order polynomials do not provide additional independent elements. It can be easily seen that $a_1$ and $b_1$ correspond to the Casimir operators of the subalgebra, that $a_1^2+b_1+b_2$ corresponds to the quadratic Casimir operator of $\mathfrak{so}(5)$, while the fourth-order invariant is given by  $$D_1+D_2+4a_1^2(b_1+b_2)+4b_1b_2+2b_1^2+2a_1^4.$$ 

\medskip
The polynomials $\{a_1,b_1,b_2\}$ Poisson-commute with each other, as well as with all other polynomials of higher degree. In addition, there are two functional relations, one involving the sum of two degree-four elements and another involving the sum of the degree-six elements: 
\begin{align}
D_1+D_2-2b_2^2 =0 \, , \qquad 
F_1+F_2-2C_1^2-b_2 D_3=0 \label{funrels1} \, .
\end{align}
From the perspective of the labelling problem, these algebraic relations imply that, as labelling operators, $D_1$ (or $D_2$) and $F_1$ (or $F_2$) would be superfluous, while either the cubic or fourth-order operator $C_1$ and $D_3$ could be used as the required  internal labelling operator, but not both simultaneously, as they cannot commute with each other. 

\medskip
\noindent The elements in (\ref{BA1}) generate a (cubic) polynomial with explicit relations 
\begin{equation*}
\begin{split}
 \{a_1, \cdot\}&=\{b_1, \cdot\}=\{b_2, \cdot\}=0,\quad \{C_1, D_1\}=-\{C_1, D_2\}=-\{C_1, D_3\}=2(F_1-F_2),\\
\{C_1, F_1\}&=\left( D_2+D_3/2\right)D_3-4 b_1 C_1^2+ 2 b_2\left( F_2-2 C_1^2 \right)+b_1^2\left( 4 D_1-6 b_2^2\right),\quad \{D_1, D_2\}=0  \\
\{C_1, F_2\}&=-\left( D_2+D_3/2\right)D_3 +4 b_1C_1^2+ 2 b_2 F_2-b_1^2\left( 4 D_1-6 b_2^2\right),\quad 
\{D_1, D_3\}=8 a_1 (F_1-F_2), \\
\{D_1, F_1\}&=4 \left( 2\left( C_1-a_1  b_2 \right) F_2 +2 b_1  C_1 D_1 -(a_1  D_2  +2 b_1 C_1   -2 a_1  b_2^2) D_3 -4( C_1^2- a_1  b_2  C_1+b_1  b_2^2)C_1\right), \\
\{D_1, F_2\}&=4\left(2\left( C_1-a_1 b_2\right)F_2+a_1 D_2 D_3-2 b_1 C_1 D_1+4 b_1 b_2^2 C_1\right),\quad 
\{D_2, D_3\}=-8 a_1(F_1-F_2), \\
\{D_2, F_1\}&=4 \left( -2\left( C_1-a_1 b_2\right) F_2-2 b_1 C_1 D_1 +(a_1 D_2+2 b_1  C_1  -2 a_1 b_2^2) D_3 +4 (C_1^2- a_1 b_2  C_1+b_1  b_2^2  )C_1\right), \\
\{D_2, F_2\}&=-4\left(2\left( C_1-a_1 b_2\right)F_2-2 b_1 C_1 D_1+a_1 D_2 D_3+4 b_1 b_2^2 C_1\right), \\
\{D_3, F_1\}&=8 \left(2 C_1^2-F_2-b_1 D_1+b_2 D_3+2 b_1b_2^2\right) C_1-2 a_1 D_3^2 +8 a_1 b_1 \left(2 C_1^2-2 b_1 D_1+3 b_1 b_2^2\right), \\
\{D_3, F_2\}&=-8 C_1 \left(F_2 -b_1  D_1 +2 b_1  b_2^2\right)+2 a_1 \left(D_3^2-4 b_1  \left(2 C_1^2-2 b_1  D_1 +3 b_1  b_2^2\right)\right), \\
\{F_1, F_2\}&=-C_1 D_3(D_2+2 D_3) +2 a_1 b_2 \left(D_3^2-4 b_1 \left(2 C_1^2-2 b_1 D_1+3 b_1 b_2^2\right)\right)+8 b_1 C_1 \left(2 C_1^2+(b_2-2 b_1) D_1\right)\\
&\quad  +8 b_1 C_1 \left(3 b_1b_2^2-2 b_2^3\right).
\end{split}
\end{equation*}

\noindent As for the previous case, we also have an additional  element that Poisson commutes with all the generators (\ref{BA1}) of the polynomial algebra, and given by the degree-four polynomial: $D_1+D_3+4 a_1 C_1$. It is straightforward to verify that $$\{D_1+D_3+4 a_1 C_1, \cdot\}=0 \, .$$
Taking into account the two functional relations \eqref{funrels1} and the quartic element above,  we close a three-generator cubic algebra (in this case spanned by seven elements) with four central elements. Considering the new generators
\begin{align}
c_1:=a_1 \, , \quad c_2&:=b_1 \, , \quad c_3:= b_2\, , \quad  c_4=D_1+D_3+4 a_1 C_1 \, , \quad  \mathcal{A}:=C_1 \, ,\quad  \mathcal{B}:=D_1 \, \, ,\quad \mathcal{C}:=2(F_1 -F_2),  
\label{eq:bas}
\end{align}
the following polynomial relations are obtained:
\begin{equation}\label{eqcc1}
\begin{split}
\{\mathcal{A}, \mathcal{B}\}&=\mathcal{C} \\
\{\mathcal{A},\mathcal{C}\}&=\alpha \mathcal{A}^2+\beta \mathcal{B}^2+2\gamma \mathcal{A} \mathcal{B} +\delta \mathcal{A}+\epsilon \mathcal{B} +\zeta\\
\{\mathcal{B},\mathcal{C}\}&=\lambda \mathcal{A}^3+\mu \mathcal{A}^2+\nu \mathcal{B}^2+2\xi \mathcal{A} \mathcal{B}+\rho \mathcal{A}+\sigma \mathcal{B} +\chi
\end{split}
\end{equation}
for the values
\begin{equation}
\begin{split}
\alpha&=-\xi= 8 (4 c_1^2 - 2 c_2 - c_3), \quad \beta=6 \, , \quad \gamma=-\nu=16 c_1 \, , \quad \delta=-\sigma=- 
16  c_1 (c_3^2 + c_4),\\
 \epsilon&=4 (4 c_2^2 - c_3^2 - 2 c_4), \quad \zeta=4 c_3^2( c_4-6 c_2^2) + 2c_4^2, \quad \lambda=-32, \quad \mu=96c_1 c_3,\\
 \rho&=16 c_3 (4 ( c_1^2 - c_2) c_3 - c_4), \quad \chi=  -16 c_1 c_3^2 c_4,
\end{split}
\end{equation}
where $c_i$ ($i=1,2,3,4$) are central elements.

\medskip
\noindent For this choice of generators, the degree of each polynomial is:
\begin{equation}
\begin{split}
{\rm deg}(c_1)=1,\;  {\rm deg}(c_2)={\rm deg}(c_3)=2,\; {\rm deg}(c_4)=4,\; {\rm deg}(\mathcal{A})=3,\;  {\rm deg}(\mathcal{B})=4,\; {\rm deg}(\mathcal{C})=6
\end{split}
\end{equation}
and the Jacobi identity \eqref{jac} is trivially satisfied. The cubic algebra possesses a quartic Casimir given by 
\begin{equation}\label{ccas2}
\begin{split}
\mathcal{K} &=\mathcal{C}^2+\frac{\lambda }{2}\mathcal{A}^4+\frac{2}{3} \mu \mathcal{A}^3 -\frac{2}{3}\beta \mathcal{B}^3-2 \alpha \mathcal{A}^2 \mathcal{B}-2\gamma \mathcal{A} \mathcal{B}^2+\rho \mathcal{A}^2-\epsilon \mathcal{B}^2
-2\delta \mathcal{A} \mathcal{B}+ 2 \chi \mathcal{A}- 2\zeta \mathcal{B}  \, .
\end{split}
\end{equation}
Using the central elements $c_i$ ($i=1,2,3,4$), the Casimir can be rewritten as 
\begin{equation}
\mathcal{K}=4 c_3^2( 8c_2^2 c_3^2-  c_4^2) \, .
\label{eq:centrals}
\end{equation}
As expected, the polynomial cubic algebra \eqref{eqcc1} can be presented uniformly in terms of a polynomial function
\begin{equation}
h(\mathcal{A},\mathcal{B})=-\frac{\lambda }{4}\mathcal{A}^4-\frac{1}{3} \mu \mathcal{A}^3 +\frac{1}{3}\beta \mathcal{B}^3+ \alpha \mathcal{A}^2 \mathcal{B}+ \gamma \mathcal{A} \mathcal{B}^2-\frac{1}{2}\rho \mathcal{A}^2+\frac{1}{2}\epsilon \mathcal{B}^2
+\delta \mathcal{A} \mathcal{B}- \chi \mathcal{A}+\zeta \mathcal{B} \, ,
\end{equation}
with the relations recovered as 
\begin{equation}
\begin{split}
\{\mathcal{A},\mathcal{B}\}&=\mathcal{C},\quad \{\mathcal{A},\mathcal{C}\}=\frac{\partial h(\mathcal{A},\mathcal{B})}{\partial \mathcal{B}},\quad \{\mathcal{B},\mathcal{C}\}=-\frac{\partial h(\mathcal{A},\mathcal{B})}{\partial \mathcal{A}}.
\end{split}
\end{equation}

\medskip
For the embedding $\mathfrak{so}(5) \supset \mathfrak{su}(2) \times \mathfrak{u}(1)$, the six labels required to label unambiguously an IR of $\mathfrak{so}(5)$ can be deduced from (\ref{eqcc1}) to construct an orthogonal basis of states: two labels correspond to the eigenvalues of the Casimir operators of $\mathfrak{so}(5)$, the Casimir operator $b_1$ of $\mathfrak{su}(2)$ and $v_3$ are used to determine the subgroup representations appearing in the decomposition (\ref{irde}), while $u_3$ distinguishes states within each $\mathfrak{su}(2)$ multiplet. As ``missing label" that separates degeneracies, any polynomial $\Phi(\mathcal{A},\mathcal{B},\mathcal{C})$ can be used. By construction, the basis is orthogonal.

\subsection{ The chain $\hat{S}(3) \supset \mathfrak{sl}(2,\mathbb{R}) \times \mathfrak{so}(2)$}
\label{sec3.3}

As final example, we consider a reduction chain related to a non-semisimple Lie algebra, namely the centrally-extended Schr{\"o}dinger algebra $\hat{S}(3)$ in $(3+1)$-dimensional space-time, which has been analyzed in detail in the context of maximal kinematical groups of the massive Schrödinger particle \cite{hag72, Nie72, BarXu81, Dob97}. The Lie algebra $\hat{S}(3)$ is given by the basis $\{J_{12}, J_{13}, J_{23},P_1, P_2, P_3,G_1,G_2,G_3,D,K,P_t,M\}$, where $\{J_{12}, J_{13}, J_{23}\}$  and $\{P_1, P_2, P_3\}$ are the generators of rotations and spatial translations respectively, $P_t$ represents the time translation generator, $\{G_1,G_2,G_3\}$ the generators related to special Galilei transformations, $D$ the generator of scale transformations, $K$ the generator of Galilean conformal transformations and $M$ is a central element. Over this basis, the table of commutator reads

\begin{center}
\begin{tabular}{c|cccccccccccc}
${[\cdot, \cdot]}$ & ${J_{12}}$ &  ${J_{13}}$ &  ${J_{23}}$ & ${P_1}$ & ${P_2}$ & ${P_3}$ & ${G_1}$ & ${G_2}$ & ${G_3}$ & ${K}$ & ${D}$ & ${P_t}$  \\	\hline
${J_{12}}$ & $0$ & $J_{23}$  & $-J_{13}$  & $P_2$ & $-P_1$ & $0$ & $G_2$ & $-G_1$ & $0$ & $0$ & $0$ & $0$ \\	 
${J_{13}}$ & $-J_{23}$ & $0$ & $J_{12}$ & $P_3$ & $0$ & $-P_1$ & $G_3$ & $0$ & $-G_1$ & $0$ & $0$ & $0$  \\	 
${J_{23}}$  & $J_{13}$ & $-J_{12}$ & $0$ & $0$ & $P_3$ & $-P_2$ & $0$ & $G_3$ & $-G_2$ & $0$ & $0$ & $0$  \\	 
${P_1}$ & $-P_2$ & $-P_3$ & $0$ & $0$ & $0$ & $0$ & $M$ & $0$ & $0$ & $G_1$ & $P_1$ & $0$  \\	 
${P_2}$ & $P_1$ & $0$ & $-P_3$ & $0$ & $0$ & $0$ & $0$ & $M$ & $0$ & $G_2$ & $P_2$ & $0$  \\	 
${P_3}$ & $0$ & $P_1$ & $P_2$ & $0$ & $0$ & $0$ & $0$ & $0$ & $M$ & $G_3$ & $P_3$ & $0$  \\	 
${G_1}$ & $-G_2$ & $-G_3$ & $0$ & $-M$ & $0$ & $0$ & $0$ & $0$ & $0$ & $0$ & $-G_1$ & $-P_1$\\	 
${G_2}$ & $G_1$ & $0$ & $-G_3$ & $0$ & $-M$ & $0$ & $0$ & $0$ & $0$ & $0$ & $-G_2$ & $-P_2$ \\	 
${G_3}$ & $0$ & $G_1$ & $G_2$ & $0$ & $0$ & $-M$ & $0$ & $0$ & $0$ & $0$ & $-G_3$ & $-P_3$ \\	 
${K}$ & $0$ & $0$ & $0$ & $-G_1$ & $-G_2$ & $-G_3$ & $0$ & $0$ & $0$ & $0$ & $-2 K$ & $-D$  \\	 
${D}$ & $0$ & $0$ & $0$ & $-P_1$ & $-P_2$ & $-P_3$ & $G_1$ & $G_2$ & $G_3$ & $2 K$  & $0$ & $-2 P_t$ \\	 
${P_t}$ & $0$ & $0$ & $0$ & $0$ & $0$ & $0$ & $P_1$ & $P_2$ & $P_3$ & $D$ & $2 P_t$ & $0$  \\	\hline
\end{tabular}
\end{center}
We analyze the commutant related to the subalgebra $\mathfrak{sl}(2,\mathbb{R}) \times \mathfrak{so}(2)$ involved in the Lie algebra-subalgebra chain $\hat{S}(3) \supset \mathfrak{sl}(2,\mathbb{R}) \times \mathfrak{so}(2)$.
In this case, we consider the following coordinates: $$\boldsymbol{x}:=\{x_1, x_2, x_3, x_4, x_5, x_6,x_7,x_8,x_9,x_{10},x_{11},x_{12},x_{13}\} \equiv\{j_{12}, j_{13}, j_{23},p_1,p_2,p_3,g_1,g_2,g_3, p_t, d, k,m\}$$ where $m$ refers to the central element, and the coordinates associated to the subalgebra are $$\boldsymbol{x}':=\{x_1, x_{10}, x_{11}\}\cup \{x_{13}\} \equiv\{j_{12}, p_t, d\} \cup \{m\} \, .$$ The system (\ref{mlpa}) is given in this case by 
\begin{equation}
\{j_{12}, p^{(n)}(\boldsymbol{x})\}=0,\quad 
\{p_t, \,p^{(n)}(\boldsymbol{x})\}=0,\quad 
\{d, p^{(n)}(\boldsymbol{x})\}=0,
\label{pdees}
\end{equation}
and possesses  nine (functionally) independent solutions, hence we expect a polynomial algebra generated by at least this number of operators. Up to degree five, the following fourteen polynomials satisfy the system (\ref{pdees}) and are linearly independent 

\begin{equation}\label{BS3}
\begin{split}
p_1^{(1)}=&\;j_{12},\quad p_2^{(2)}=j_{13}^2+j_{23}^2,\quad p_3^{(2)}= p_1 g_2-p_2 g_1,\quad p_4^{(2)}=k p_t-\frac{1}{4}d^2,\\
p_5^{(3)}=&\;j_{13}(p_1 g_3-p_3 g_1)+j_{23}(p_2 g_3-p_3 g_2),\quad p_6^{(3)}
=j_{13}(p_2 g_3-p_3 g_2)+j_{23}(p_3 g_1-p_1 g_3),\\
p_7^{(3)}=&\;k(p_1^2+p_2^2)+p_t(g_1^2+g_2^2)-d(g_1 p_1+g_2 p_2),\quad p_8^{(3)}=k p_3^2+p_t g_3^2-d p_3 g_3,\\
p_9^{(4)}=&\;j_{13}(k p_1 p_3+p_t g_1 g_3)+j_{23}(k p_2 p_3+p_t g_2 g_3)-\frac{d}{2}\bigl(j_{13}(p_1 g_3+ p_3 g_1)+j_{23}(p_2 g_3+p_3 g_2)\bigl),\\
p_{10}^{(4)}=&\;j_{13}(k p_2 p_3+p_t g_2 g_3)-j_{23}(k p_1 p_3+p_t g_1 g_3)-\frac{d}{2}\bigl(j_{13}(p_2 g_3-p_3 g_2)-j_{23}(p_1 g_3+p_3 g_1)\bigl),\\
p_{11}^{(4)}=&\;p_3^2 (g_1^2+g_2^2)+g_3^2(p_1^2+p_2^2)-2 p_3g_3(p_1 g_1+p_2 g_2),\\
p_{12}^{(5)}=&\;k (j_{13} p_1+j_{23}p_2)^2+p_t (j_{13}g_1+j_{23}g_2)^2-d(j_{13} p_1+j_{23}p_2)(j_{13}g_1+j_{23}g_2),\\
p_{13}^{(5)}=&\;p_t(j_{13}g_2-j_{23}g_1)(j_{13}g_1+j_{23}g_2)-k(j_{23}p_1-j_{13}p_2)(j_{13}p_1+j_{23}p_2)+ d g_1j_{13}j_{23}p_1\\
& + \frac{d g_1p_2}{2} (j_{23}^2-j_{13}^2)\\
p_{14}^{(5)}=&\;\bigl(\frac{d}{2}(g_1^2+g_2^2)-k(p_1 g_1+p_2 g_2)\bigl)p_3^2+\bigl(p_t(p_1 g_1+p_2 g_2)-\frac{d}{2}(p_1^2+p_2^2)\bigl)g_3^2+k(p_1^2+p_2^2)g_3 p_3\\
& -p_t(g_1^2+g_2^2)g_3 p_3.
\end{split}
\end{equation}
To these elements, the central charge $m$ must be added. We rename the preceding elements following the notational conventions adopted in Section \ref{sec2.1}:
\begin{equation}\label{NB}
\begin{split}
 a_1 =&\; m,\quad a_2=p_{1}^{(1)},\quad B_1=p_{2}^{(2)},\quad B_2=p_{3}^{(2)},\quad b_1=p_{4}^{(2)},\quad C_1=p_{5}^{(3)},\quad C_2 = p_{6}^{(3)},\quad C_3=p_{7}^{(3)}\\
 C_4 =&\; p_{8}^{(3)},\quad D_1=p_{9}^{(4)},\quad D_2=p_{10}^{(4)},\quad D_3=p_{11}^{(4)},\quad 
 E_1= p_{12}^{(5)},\quad E_2=p_{13}^{(5)},\quad E_3=p_{14}^{(5)}.
\end{split}
\end{equation}
There exist some algebraic relations between these elements, such as 
\begin{equation}
C_1^2+C_2^2-B_1D_3=0 \, , \quad C_2 D_1-C_1 D_2+B_1 B_2 C_4=0 \, , \quad C_1 D_1+C_2 D_2-B_1 E_3=0 \label{f71}\, .
\end{equation}
Observe however that this does not allow us to discard any of the involved generators, as they are not polynomial relations, but rational functions of the elements, and thus do not belong to $\mathsf{Pol}(\hat{S}(3)^{\ast})$. The functions $a_2$ and $b_1$ are clearly the Casimir invariants of the subalgebra $\mathfrak{sl}(2,\mathbb{R}) \times \mathfrak{so}(2)$, while the Casimir invariants of the $\hat{S}(3)$ are given by the central element $a_1$ and the third- and fourth-order polynomials given by 
\begin{equation}
\begin{split}
\frac{m}{2}\left(a_2^2+B_1-4b_1\right)-a_2B_2-C_1+C_3+C_4,\\
a_1^2\left(a_2^2+B_1\right)-2a_1a_2B_2+B_2^2-2a_1C_1+D_3. 
\end{split}
\end{equation} 
Added to the five Casimirs, any other four (functionally) independent polynomials of (\ref{BS3}) would provide a complete set of labelling operators. Among these four additional polynomials, at most two are expected to commute \cite{cam09}. 
The elements $\{a_1\, ,  \, a_2, \, b_1\}$ Poisson commute with all the polynomials in (\ref{NB}), and can thus be considered as central elements of the polynomial algebra generated by $\{B_1, B_2, C_1, C_2, C_3, C_4, D_1,D_2,D_3, E_1,E_2,E_3 \}$.
Besides $\{a_1, \cdot\}=\{a_2, \cdot\}=\{b_1, \cdot\}=0$, we get the following Poisson brackets up to degree six
\begin{equation}\label{KL1}
\begin{split}
\{B_1,B_2\}= &-2 C_2,\quad  \{B_1,C_1\}=2 a_2 C_2,\quad \{B_1,C_2\}=2(B_1 B_2-a_2 C_1),\quad \{B_1,C_3\}= 4 D_1,\\
\{B_1,C_4\}= & -4D_1,\quad \{B_2, C_1\}=a_1 C_2,\quad \{B_2, C_2\}=D_3-a_1 C_1,\quad \{B_2, C_3\}=\{B_2, C_4\}=0,\\
\{B_1,D_1\}=& 2(B_1 C_4-E_1+a_2 D_2),\quad \{C_1,C_2\}=a_1 B_1 B_2-a_2 D_3,\quad \{B_1,D_2\}=-2(E_2+a_2 D_1),\\
\{C_1,C_3\}= & -\{C_1,C_4\}=2 (E_3+a_1 D_1),\quad \{B_1,D_3\}=4 B_2 C_2,\quad  \{B_2, D_1\}=-B_2 C_4+a_1 D_2,\\
\{C_2,C_3\}= & 2(-B_2 C_4+a_1 D_2),\quad \{B_2, D_2\}=E_3-a_1 D_1,\quad \{C_2,C_4\}=-2(-B_2 C_4+a_1 D_2),\\
\{B_2, D_3\}=&0,\quad \{C_3,C_4\}=-2 E_3,\quad \{B_1,E_1\}=4 (B_1 D_1+a_2 E_2),\quad \{B_2, E_1\}=2(a_1 E_2-B_2 D_1),\\
\{B_1,E_2\}=&2B_1D_2+2a_2 (B_1 C_3-2  E_1),\quad \{B_1,E_3\}=4 B_2 D_2-2C_1(C_3-C_4),\quad \{B_2, E_3\}=0,\\
\{B_2, E_2\}=&C_1 C_3-2 B_2 D_2 +a_1(B_1 C_3-2 E_1),\quad \{C_2,D_1\}=C_2 C_4-a_1 E_2+a_2 E_3,\\
\{C_1,D_1\}=&2 B_2 D_2-C_1 \left(C_3-C_4\right)+a_1 \left(B_1 C_4-E_1\right)+a_2 B_2 C_4,\quad  \{C_1,D_3\}=2 a_1 B_2 C_2,\\
\{C_1,D_2\}=&-2 B_2 D_1-C_2(C_3-C_4)-a_1 E_2-a_2 E_3,\quad \{C_2,D_3\}=2 B_2(D_3-a_1 C_1),\\ 
\{C_2,D_2\}=&-C_1C_4+a_1(E_1-B_1(C_3-C_4))+a_2 B_2 C_4,\quad \{C_3,D_3\}=-4 a_1 E_3,\\
\{C_3,D_1\}=&2 B_2 D_2-C_3(C_1+2C_4)+2 b_1 (D_3+a_1 C_1),\quad \{C_3,D_2\}=2 (a_1 b_1 C_2- B_2 D_1)-C_2C_3,\\
\{C_4,D_1\}=&(C_1+2C_3)C_4-2 b_1 (D_3+a_1 C_1),\quad \{C_4,D_2\}=C_2(C_4-2 a_1 b_1),\quad \{C_4,D_3\}=4 a_1 E_3.
\end{split}
\end{equation}
The remaining relations give rise to polynomials up to degree nine: 
\begin{equation}\label{KL2}
\begin{split}
 \{C_1,E_1\}=&2(B_2 E_2+(a_1 B_1+a_2 B_2+C_1)D_1),\\
 \{C_1,E_2\}=&B_2 \left(B_1(C_3-C_4)-2 E_1\right)+2 C_1 D_2+a_1 B_1 D_2+ a_2 (2B_2D_2- C_1 C_3),\\
\{C_1,E_3\}=&2 B_2^2 C_4-(C_3-C_4)D_3+a_1(2 B_2D_2-C_1(C_3-C_4)),\\
\{C_2,E_1\}=&2(C_2 D_1-B_2 E_1-a_2(B_2 D_2-C_1 C_3)),\\
\{C_2,E_2\}=&B_1 E_3-2 B_2 E_2-2 C_1 D_1+a_1 B_1 D_1+a_2 (C_2 C_3+2 B_2 D_1),\\
\{C_2,E_3\}=&2 B_2 E_3-a_1(C_2(C_3-C_4)+2 B_2 D_1),\\
\{C_3,E_1\}=&2 B_2 E_2-4 C_3 D_1-4 b_1 B_2 C_2,\\
\{C_3,E_2\}=&B_2(B_1 C_3-2 E_1) -2 C_3 D_2+4 b_1B_2 C_1,\\
\{C_3,E_3\}=&2 B_2^2C_4-C_3D_3-2a_1(C_3 C_4-2 b_1 D_3),\\
\{C_4,E_1\}=&2 (C_1+2C_3)D_1+4 b_1 B_2 C_2,\\
\{C_4,E_2\}=&2(C_1+C_3)D_2-B_2(B_1 C_4+4b_1 C_1),\\
\{C_4,E_3\}=&C_4 D_3+2 a_1 (C_3 C_4-2 b_1 D_3),\\
\{D_1,D_2\}=&-C_3 D_2-B_1 B_2 C_4-a_2 (C_3 C_4-b_1 D_3)+b_1 B_2(2 C_1+a_1  B_1),\\
\{D_1,D_3\}=&-2B_2^2C_4-2a_1(B_2 D_2-C_1(C_3-C_4)),\\
\{D_2,D_3\}=&2 B_2 E_3+2a_1(B_2 D_1+C_2(C_3-C_4)),\\
\{D_1,E_1\}=&(C_1+2C_3)E_1-2 b_1 (B_1(B_2^2+a_1 C_1)+C_1(C_1+a_2 B_2)),\\
\{D_1,E_2\}=&(C_1+C_3)E_2-(B_1 B_2+a_2 C_3)D_1-b_1(2 C_1+a_1 B_1+2a_2 B_2)C_2,\\
\{D_1,E_3\}=&-(C_3-C_4)(E_3-a_1D_1)-B_2 C_2(C_4-2 a_1 b_1),\\
\{D_2,E_1\}=&(C_1+2C_3)E_2+B_1 B_2 D_1+2a_2(C_3 D_1+b_1 B_2 C_2)-2 b_1 C_1 C_2,\\
\{D_2,E_2\}=&B_1 C_3^2-C_1 E_1+C_3(B_1 C_1-E_1+a_2 D_2)-b_1\bigl(B_1(2 B_2^2+D_3+a_1 C_1)\bigl)-2b_1C_1\times\\
& (C_1-a_2 B_2),\\
\{D_2,E_3\}=&B_2(C_4(C_1+C_3+C_4)-2 b_1 D_3)+a_1((C_3-C_4)D_2-2 b_1 B_2 C_1),\\
\{D_3,E_1\}=&4 (B_2^2+a_1 C_1)D_1,\\
\{D_3,E_2\}=&4(B_2^2+a_1 C_1)D_2-2 B_2(C_1 C_3 +a_1 B_1 C_4),\\
\{D_3,E_3\}=&2a_1(2 B_2^2 C_4-(C_3-C_4)D_3),\\
\{E_1,E_2\}=&-(B_1 B_2+2 a_2 C_3)E_1+2 b_1 B_1 B_2(C_1+a_1 B_1+2 a_2 B_2),\\
\{E_1,E_3\}=&(B_2 D_2-C_1C_3)(C_1+2(C_3-C_4))-B_2(C_2D_1-4 b_1 B_2C_1)-a_1(2 C_4E_1-4 b_1 C_1^2)\\
\{E_2,E_3\}=&2 B_2 C_4 D_1-C_2 C_3 (C_1+C_3-C_4)-2a_1(C_4 E_2-2 b_1 C_1 C_2)+4 b_1 B_2^2C_2.
\end{split}
\end{equation}

\noindent This shows that the fifteen polynomials constructed from the commutant close a cubic polynomial algebra, although it is not excluded that additional algebraic dependency relations in higher order imply some simplification of the structure. The interesting fact is that the polynomial algebra (\ref{KL1})-(\ref{KL2}) contains some quadratic polynomial subalgebras. Consider for example the polynomials $\{a_1, a_2,B_1, B_2, C_1, C_2, D_3\}$.  The centre is given by $a_1,a_2$, while the remaining generators close a quadratic algebra with the following Poisson brackets
\begin{equation}
\begin{split}
\{B_1,B_2\}& =-2 C_2,\quad  \{B_1,C_1\}=2 a_2 C_2,\quad  \{B_1,C_2\}=2(B_1 B_2-a_2 C_1),\quad \{B_1,D_3\}=4 B_2 C_2,\\
\{B_2, C_1\}&=a_1 C_2,\quad \{B_2, C_2\}=D_3-a_1 C_1,\quad \{B_2,D_3\}=0,\quad  \{C_1,C_2\}=a_1 B_1 B_2-a_2 D_3,\\ 
\{C_1,D_3\}&=2 a_1 B_2 C_2,\quad \{C_2,D_3\}=2 B_2(D_3-a_1 C_1) .
\end{split}
\end{equation}

Besides the already mentioned Casimir operators, in order to construct an orthogonal set of labelling operators, three operators are required. As internal label the generators $D$ of the scale transformations can be considered, while for the two necessary commuting operators to separate degeneracies, the simplest choice is given by $B_2$ and an arbitrary polynomial $\Phi(C_3,C_4,D_3,E_3)$ (see (\ref{KL1})-(\ref{KL2})).

\section{The Elliott chain and its associated cubic commutator algebra}
\label{Sec4}

\noindent In this section, we revisit the case of the Elliott model $\mathfrak{su}(3) \supset \mathfrak{so}(3)$ analyzed in Section \ref{sec3.1} from commutator algebra point of view in terms of the enveloping algebra $\mathcal{U}(\mathfrak{su}(3))$, in order to illustrate the technical differences between the commutative and noncommutative approaches. As mentioned, to obtain the corresponding elements in the enveloping algebra $\mathcal{U}(\mathfrak{su}(3))$, we have to symmetrize the polynomials in \eqref{eq:basis} using (\ref{symmap}). The computations and simplifications have been done relying on the symbolic computer algebra software  \textsf{Mathematica}\textsuperscript{\tiny{\textregistered}}, specifically on the package \textsf{NCAlgebra} \cite{NCA}.

Taking into account the results obtained in the commutative setting (see equations \eqref{eq:basis}-(\ref{eqc1})), it suffices to symmetrize the homogeneous polynomials $b_1=c_1$, $b_2=c_2$, $C_1$, $C_2$ and $D_1$, as the symmetrization of $F_1$ follows from the properties of the commutator (see equation (\ref{polrelq})). Application of the symmetrization operator (\ref{symmap}) to these polynomials leads to 
\begin{equation}
\begin{split}
\mathsf{b}_1=&\; \textsf{S}_2(L_1, L_1)+\textsf{S}_2(L_2, L_2)+\textsf{S}_2(L_3, L_3)\\
\mathsf{b}_2=&\; \textsf{S}_2(T_{11}, T_{11})+\textsf{S}_2(T_{12}, T_{12})+\textsf{S}_2(T_{13}, T_{13})+\textsf{S}_2(T_{11}, T_{22})+\textsf{S}_2(T_{22}, T_{22})+\textsf{S}_2(T_{23}, T_{23})\\ 
\mathsf{C}_1=&\;  \textsf{S}_3(L_3,L_3, T_{11}) +\textsf{S}_3(L_3,L_3, T_{22})-\textsf{S}_3(L_1,L_1, T_{11})  - 2\left( \textsf{S}_3(L_1,L_2, T_{12}) +\textsf{S}_3(L_1,L_3, T_{13})\right)\\
&  -\textsf{S}_3(L_2,L_2, T_{22})-2 \textsf{S}_3(L_2,L_3, T_{23}) \\
\mathsf{C}_2 =&\; \textsf{S}_3(T_{12}, T_{13}, T_{23})+\frac{1}{2} \left(\textsf{S}_3(T_{11}, T_{12},T_{12}) -\textsf{S}_3(T_{11},  T_{22}, T_{22}) - \textsf{S}_3(T_{11}, T_{23}, T_{23})\right)\\
& +\frac{1}{2}\left(\textsf{S}_3(T_{22}, T_{12},T_{12}) - \textsf{S}_3(T_{22},T_{13}, T_{13})-\textsf{S}_3(T_{22}, T_{11}, T_{11})\right)\\
\mathsf{D}_1=&\; \textsf{S}_4(L_1,  L_1, T_{11}, T_{11}) +\textsf{S}_4(L_1, L_1, T_{12}, T_{12}) + \textsf{S}_4(L_1, L_1, T_{13}, T_{13}) + 2 \textsf{S}_4(L_2, L_3, T_{12}, T_{13})\\
& + 2\left( \textsf{S}_4(L_1,  L_2, T_{12}, T_{11}) + \textsf{S}_4(L_1,  L_2, T_{12},  T_{22})+  \textsf{S}_4(L_1,L_2, T_{13}, T_{23})+\textsf{S}_4(L_2, L_3, T_{11}, T_{23})\right)\\
& +2 \left(\textsf{S}_4(L_1,L_3, T_{12}, T_{23})- \textsf{S}_4(L_1, L_3, T_{13}, T_{22})\right)  + \textsf{S}_4(L_2,L_2, T_{12},T_{12}) + \textsf{S}_4(L_2,L_2, T_{22},T_{22})\\
& +  \textsf{S}_4(L_2,L_2, T_{23},T_{23})  +   \textsf{S}_4(L_3,L_3, T_{13},T_{13})  + \textsf{S}_4(L_3,L_3, T_{11},T_{11})+\textsf{S}_4(L_3,L_3, T_{22},T_{22})\\
& +2\textsf{S}_4(L_3,L_3, T_{11},T_{22}) + \textsf{S}_4(L_3,L_3, T_{23},T_{23}).
\end{split}
\end{equation}
Expanding each of the terms and reordering terms with respect to the ordered basis (\ref{subTij}) we obtain the following elements in the enveloping algebra $\mathcal{U}\left(\mathfrak{su}(3)\right)$
\begin{equation}
\begin{split}
\mathsf{b}_1=&\;L_1^2 + L_2^2 + L_3^2 \\
\mathsf{b}_2=&\;T_{11}^2 + T_{12}^2 + T_{13}^2 + T_{11} T_{22} + T_{22}^2 + T_{23}^2\\
\mathsf{C}_1=&\;L_3^2 (T_{11} + T_{22}) - 2 L_1 (L_2 T_{12} + L_3 T_{13}) - 2 L_2 L_3 T_{23} + {\rm i} (L_1 T_{23}-L_2 T_{13}+L_3 T_{12})\\
& -L_1^2 T_{11}- L_2^2 T_{22}\\
\mathsf{C}_2=&\; T_{12} T_{13} T_{23}+\frac{1}{2} \bigl(T_{11} (T_{12}^2 - T_{22}^2 - T_{23}^2)-T_{22}(T_{11}^2  - T_{12}^2  + T_{13}^2) \bigl)+2 T_{11}-T_{22}\\
& -\frac{{\rm i}}{2}(L_1 T_{23}-L_2 T_{23}+5 L_3T_{12})\\
\mathsf{D}_1=&\;L_1^2 (T_{11}^2 + T_{12}^2 + T_{13}^2) + 
2 L_1 \bigl(L_2 (T_{12} (T_{11} + T_{22}) +  T_{13} T_{23}) +L_3(T_{12} T_{23}- T_{13} T_{22} )\bigl)\\
& + 2 L_2 L_3 (T_{12} T_{13} - T_{11} T_{23}) + L_2^2 (T_{12}^2 + T_{22}^2 + T_{23}^2)+ L_3^2 (T_{13}^2 + (T_{11} + T_{22})^2 + T_{23}^2)\\
& +{\rm i} \left( L_2(T_{12}T_{23}-T_{13}T_{22})-L_1(L_2 L_3-T_{11}T_{23}+T_{12}T_{13})-L_3(T_{11}T_{12}+T_{12}T_{22}+T_{13}T_{23})\right)\\
& -\frac{1}{2}(T_{11}^2+T_{12}^2+T_{13}^2+T_{11}T_{22}+T_{22}^2+T_{23}^2)+\frac{1}{6}(13 L_1^2+13 L_2^2-11 L_3^2).
\end{split}
\end{equation}
Comparing these expressions with those obtained in the Poisson setting, it can be easily seen that the leading-order term appearing in each polynomial coincides with the homogeneous polynomial constructed in the commutative setting, as expected from \eqref{orba}. Hence, the representative in the enveloping algebra is obtained with a leading-order term that equals that of its classical counterpart, to which the unavoidable lower-order terms due to the noncommutativity of the generators must be added. In order to construct the analogue of the polynomial Poisson algebra \eqref{eqc1} in $\mathcal{U}\left(\mathfrak{su}(3)\right)$, we consider the operators
\begin{equation}\label{eq:basiss}
\begin{split}
 \mathsf{c}_1&=\mathsf{b}_1\, , \quad \mathsf{c}_2=\mathsf{b}_2 \, , \quad \mathsf{c}_3= \frac{1}{2}\mathsf{C}_1+ \mathsf{C}_2 \, , \quad   \mathsf{A}=\frac{1}{2}\mathsf{C}_1- \mathsf{C}_2\, ,\quad   \mathsf{B}= \mathsf{D}_1 \, ,\quad  \mathsf{C}=[ \mathsf{A}, \mathsf{B}], 
\end{split}
\end{equation}
in terms of which the following cubic commutator algebra results:
\begin{equation}\label{eq1}
\begin{split}
 [\mathsf{A},\mathsf{B}]&=\mathsf{C} \\
 [\mathsf{A},\mathsf{C}]&= \alpha \mathsf{A}^2  + \beta \mathsf{B}^2 + \delta \mathsf{A}  + \epsilon \mathsf{B} + \zeta\\
[\mathsf{B},\mathsf{C}]&= \lambda \mathsf{A}^3 + \mu \mathsf{A}^2  + \xi \{\mathsf{A},\mathsf{B}\} + \rho \mathsf{A} + \sigma \mathsf{B} + \chi 
\end{split}
\end{equation}
subjected to the constraints 
\begin{equation} \label{cf1}
\begin{split}
 \alpha&=-\xi=-8 (3 \mathsf{c}_1+\mathsf{c}_2-18) \, \quad \delta=-\sigma=- 16( \mathsf{c}_2 -9)\mathsf{c}_3 \, , \quad \epsilon = -( 16  \mathsf{c}_1 \mathsf{c}_2+152 \mathsf{c}_1+24 \mathsf{c}_2-144) \\
\zeta &=8 \mathsf{c}_3^2 (3 \mathsf{c}_1 - \mathsf{c}_2)+24 \mathsf{c}_1^3+\frac{128}{3}\mathsf{c}_1^2 \mathsf{c}_2+24 \mathsf{c}_1 \mathsf{c}_2^2-\frac{358}{3}\mathsf{c}_1^2-18 \mathsf{c}_2^2-68 \mathsf{c}_1 \mathsf{c}_2-24 \mathsf{c}_1+72 \mathsf{c}_2, \quad \beta=24,\\
\lambda&=-32 \, ,\quad \mu=-48 \mathsf{c}_3,\quad \rho=-16 \mathsf{c}_1 (\mathsf{c}_1+\mathsf{c}_2)^2+168 \mathsf{c}_1^2+24 \mathsf{c}_2^2+\frac{544}{3}\mathsf{c}_1 \mathsf{c}_2+48 \mathsf{c}_1-144 \mathsf{c}_2, \\
\chi&=16 \mathsf{c}_3\left(\mathsf{c}_1^3-\mathsf{c}_1 \mathsf{c}_2^2+\mathsf{c}_3^2-4 \mathsf{c}_1^2+\frac{3}{2}\mathsf{c}_2^2+\frac{23}{6}\mathsf{c}_1\mathsf{c}_2+\frac{3}{2}\mathsf{c}_1-\frac{9}{2}\mathsf{c}_2 \right).
\end{split}
\end{equation}
As before, the operators $\mathsf{c}_i$ play the role of central elements for $i=1,2,3$. As mentioned previously, this example illustrates explicitly that the structure constants involve lower-order correction terms in comparison with those obtained for the analytical approach (see \eqref{clst1}). As expected, the Jacobi identity
\begin{equation}
[\mathsf{A}, [\mathsf{B},\mathsf{C}]] + [\mathsf{C},[\mathsf{A},\mathsf{B}]] + [\mathsf{B},[\mathsf{C},\mathsf{A}]] =0 
\end{equation}
is trivially satisfied. The quartic Casimir \eqref{ccas} of the polynomial cubic algebra gives rise by symmetrization to the following operator 
\begin{equation}\label{cas}
\begin{split}
K &= \mathsf{C}^2+\frac{\lambda }{2}\mathsf{A}^4+\frac{2}{3} \mu\mathsf{A}^3 -\frac{2}{3}\beta\mathsf{B}^3-(2 \alpha-\beta \lambda) \mathsf{A}\mathsf{B}\mathsf{A}+\left(\rho+\frac{1}{2}    (\alpha  \beta+\epsilon) \lambda   \right)\mathsf{A}^2-\left(\epsilon-\frac{1}{3}(\beta \lambda-4 \alpha)\beta\right)\mathsf{B}^2 \\
& +\left(\frac{1}{3}\beta\mu-\delta\right)\left(\mathsf{A}\mathsf{B}+\mathsf{B}\mathsf{A}\right)+ \left(2 \chi+\frac{1}{2}\beta \delta\lambda+\frac{1}{3} \epsilon \mu\right)\mathsf{A}-\left( 2\zeta +\frac{1}{3}(3 \alpha-\beta \lambda)\epsilon-\frac{1}{3}\beta \rho\right)\mathsf{B} .
\end{split}
\end{equation}
To obtain the Casimir in terms of the three generators of the cubic algebra (\ref{eq1}) it suffices to insert the values of \eqref{cf1} into \eqref{cas}. It is straigthforward to verify that the Casimir invariant \eqref{cas}, which is of degree twelve in the underlying generators,  collapses to the Casimir invariant \eqref{ccas} if we drop the lower-order terms. We also notice that the term $\mathsf{A}\mathsf{B}\mathsf{A}$ can be re-expressed as $\mathsf{A}^2\mathsf{B}-\mathsf{A}\mathsf{C}$, taking into account $\eqref{eq1}$.

\medskip
In direct application of the symmetrization procedure (\ref{symm}), the operators $\mathsf{b}_1+2\mathsf{b}_2$, $\mathsf{C}_1+2\mathsf{C}_2$, $\mathsf{b}_1$, $L_1$ and a polynomial $\Phi\left( \mathsf{A},\mathsf{B},\mathsf{C}\right)$ considered in the analytical frame are independent and commute with each other, implying that we can find a basis of eigenvectors for any IR $\Gamma$ such that these operators are diagonal and constitute an orthogonal basis. It should be taken into account that the classical bases used in the literature, such as the Bargmann--Moshinsky, Elliot and stretched and anti-stretched bases are not orthogonal \cite{mos63,mpsw}, with a first orthogonal basis given in \cite{jud74}. The choice above constitutes a generalization, allowing us to adapt the values of the labelling problem to specific observables that may be of interest in the model.

\section{Conclusions}
\label{Sec5}

\noindent In this paper we have shown how polynomial algebras arise naturally from the analysis of reduction chains $\mathfrak{g} \supset \mathfrak{g}'$ of Lie algebras, in particular from the use of the commutant associated to the subalgebra $\mathfrak{g}'$, along the line of the recent papers \cite{cam21,cor21, cam22, cam22acta}. The construction presented in this work has been mainly developed, with the idea of showing the effectiveness of this approach, in the commutative setting. We have hence focused on the Lie-Poisson bracket constructed from the Lie algebra $\mathfrak{g}$ and inhertited by its dual $\mathfrak{g}^*$. A systematic procedure to compute homogeneous polynomials in $\textsf{Pol}(\mathfrak{g}^*)$ commuting with the subalgebra generators $\boldsymbol{x}'$, degree by degree, has been proposed. It has been shown how these polynomials in the commutant close to give polynomial Poisson algebras with respect to the corresponding Lie-Poisson bracket.
As explicit examples, we have inspected three Lie algebra-subalgebra chains, namely the Elliott chain $\mathfrak{su}(3) \supset \mathfrak{so}(3)$,  the chain $\mathfrak{so}(5) \supset \mathfrak{su}(2) \times \mathfrak{u}(1)$ related to the Seniority model, and, as an illustration on how the procedure remains valid for reduction chains that do not necessarily involve semisimple Lie algebras, the reduction  $\hat{S}(3) \supset \mathfrak{sl}(2,\mathbb{R}) \times \mathfrak{so}(2)$ involving the centrally-extended Schr\"{o}dinger algebra in $(3+1)$-dimensional space-time. For the first two cases, both related to models appearing in nuclear physics, we have used a procedure (based on the construction of certain combinations of polynomials commuting with all the generators) that allowed us to obtain specific types of (three-generator) cubic polynomial Poisson algebras that have previously been observed in other contexts, such as superintegrable systems \cite{bon94, das01, mar09, mar10}.  As a consequence of these results, it may be reasonable to expect that in this context these polynomial structures are likely to play a significant role. 

The case related to the Schr\"{o}dinger algebra leads us to a polynomial cubic structure too, but now much more intricated, due to the 
non-semisimplicity of $\hat{S}(3)$ and the high number of polynomials in the commutant of the subalgebra $\mathfrak{sl}(2,\mathbb{R}) \times \mathfrak{so}(2)$. We have shown how the construction based on this commutant leads to a cubic polynomial Poisson algebra closed by twelve polynomials up to degree five, plus three additional central elements. Interestingly enough, this polynomial algebra contains a closed quadratic subalgebra involving five of the twelve generators plus two of the three central elements. It would be interesting to shed some further light on these type of substructures, such as their classification, and analyze whether they play some specific role in the characterization of the complete commutant.

We have finally shown how the corresponding polynomial algebras in the enveloping algebra $\mathcal{U}(\mathfrak{g})$ can be recovered from their classical counterpart in $\textsf{Pol}(\mathfrak{g}^*)$ by means of a symmetrization procedure. From this perspective, we have explicitly shown, taking as a guiding example the Elliott chain, how the relations obtained in the noncommutative setting equal the ones obtained in the commutative one at the leading-order, and that the main differences are due to the presence of lower-order correction terms. The polynomial cubic structure obtained for the chain $\mathfrak{su}(3) \supset \mathfrak{so}(3)$, together with its associated fourth-order Casimir element, has been presented with a precise idea in mind, namely to emphasize that the analysis of these problems is much more manageable if one relies on the commutative setting, as computationally the problem is less demanding, due to the commutative variables. When polynomials arising from commutants are of high degree, in fact, the use of computer algebra software can present nontrivial limitations, such as for example an extremely high computation time due to the presence of several thousands terms in the expansion. On the other hand, dealing with the commutative setting allows to provide preliminary information on the leading-order terms appearing in the relations, and consequently on the form of the polynomial commutator algebra one expects to obtain, being it quadratic, cubic or even higher order. At this level, the real effort restricts to provide those additional correction terms appearing in the relations, but now with a well defined guess for the form of the latter. 

In the context of the labelling problem (see e.g. \cite{jud74,Pe76,cam11,Sh70}), the use of the polynomial algebra associated to the subalgebra can be of use to find appropriate labelling operators of any order, and adapted to the various types of representations of $\mathfrak{g}$ reduced with respect to $\mathfrak{g}^{\prime}$. As occasionally a labelling operator reduces to zero for some degenerate representation, a precise knowledge of the algebraic relations of the available labelling operators can help to find a suitable representative that does not trivialize on these degenerate multiplets, and from which a complete basis of eigenstates could be inferred. As there is no canonical choice of labelling operators, the knowledge of polynomial algebras generated by them potentially constitutes a tool to determine which choice of labelling operators is more appropriately adapted to a specific reduction chain, even providing some insight into the physical meaning of each labelling operator. To which extent polynomial Poisson algebras are suitable to systematize the labelling problem, and even to characterize the symmetry properties of bases of eigenstates, is currently being analyzed in further detail. We hope to report some progress in this direction in the near future.

\section*{Acknowledgement}
This work was partially supported by the Future Fellowship FT180100099 and the Discovery Project DP190101529 from the Australian Research Council. RCS  acknowledges financial support by the research grant PID2019-106802GB-I00/AEI/10.13039/501100011033 (AEI/ FEDER, UE).

\end{document}